\documentclass[aps, prx, reprint,footinbib,floatfix,superscriptaddress]{revtex4-2}
\usepackage{graphicx}
\usepackage{indentfirst}
\usepackage{braket}
\usepackage{float}
\usepackage{amsmath}
\usepackage{physics}
\usepackage{amssymb}
\usepackage{CJK}
\usepackage{esint}
\usepackage{color}
\usepackage[T1]{fontenc}
\usepackage{subfigure}
\usepackage{amsfonts}
\usepackage{footmisc}
\usepackage{scrextend}
\usepackage{multirow}
\usepackage[outdir=./]{epstopdf}
\usepackage{xcolor}
\usepackage[english]{babel}
\usepackage{url}
\usepackage{comment}
\usepackage{array}
\usepackage{subfiles}
\usepackage{tikz}
\usetikzlibrary{shapes,arrows}
\usepackage{mathrsfs}
\usepackage{mathtools}
\usepackage[mathscr]{eucal}

\usepackage[hyperfootnotes=false]{hyperref}
\usepackage{cleveref}
\definecolor{darkblue}{rgb}{0,0,0.5}
\hypersetup{
colorlinks=true,
linkcolor=black,
filecolor=blue,
citecolor=darkblue,  
urlcolor=cyan,
}

\def\be{\begin{equation}}
\def\ee{\end{equation}}
\def\ba{\begin{eqnarray}}
\def\ea{\end{eqnarray}}
\def\bal{\begin{equation}\begin{aligned}}
\def\eal{\end{aligned}\end{equation}}

\def\bp{\begin{pmatrix}}
\def\ep{\end{pmatrix}}

\def\qqquad{\qquad\qquad}

\usepackage{bm}

\newcommand{\calM}{{\cal M}}

\newcommand{\1}{^{(1)}}

\usepackage[normalem]{ulem}

\newcommand{\QZ}[1]{{{\textcolor{black}{#1}}}}

\begin{document}


\title{Entanglement-enhanced optomechanical sensor array for dark matter searches}

\author{Anthony J. Brady}
\thanks{These two authors contributed equally.}
\affiliation{
Department of Electrical and Computer Engineering, University of Arizona, Tucson, Arizona 85721, USA
}

\author{Xin Chen}
\thanks{These two authors contributed equally.}
\affiliation{
Department of Electrical and Computer Engineering, University of Arizona, Tucson, Arizona 85721, USA
}

\author{Kewen Xiao}
\affiliation{
School of Physics and Astronomy, Rochester Institute of Technology, Rochester, NY 14623, USA}

\author{Yi Xia}
\affiliation{
Department of Materials Science and Engineering, University of Arizona, Tucson, Arizona
85721, USA}
\affiliation{
J. C. Wyant College of Optical Sciences, University of Arizona, Tucson, Arizona 85721, USA}

\author{Jack Manley}
\affiliation{
J. C. Wyant College of Optical Sciences, University of Arizona, Tucson, Arizona 85721, USA}

\author{Mitul Dey Chowdhury}
\affiliation{
J. C. Wyant College of Optical Sciences, University of Arizona, Tucson, Arizona 85721, USA}

\author{Zhen Liu} 
\affiliation{
School of Physics and Astronomy, University of Minnesota, Minneapolis, MN 55455, USA
}

\author{Roni Harnik} 
\affiliation{
Theoretical Physics Division, Fermi National Accelerator Laboratory, P.O. Box 500, Batavia, Illinois 60510, USA}

\author{Dalziel J. Wilson}
\affiliation{
J. C. Wyant College of Optical Sciences, University of Arizona, Tucson, Arizona 85721, USA}

\author{Zheshen Zhang}%
\affiliation{
Department of Materials Science and Engineering, University of Arizona, Tucson, Arizona
85721, USA}
\affiliation{
J. C. Wyant College of Optical Sciences, University of Arizona, Tucson, Arizona 85721, USA}
\affiliation{Department of Electrical Engineering and Computer Science, University of Michigan, Ann Arbor, MI 48109, USA}

\author{Quntao Zhuang}%
\email{qzhuang@usc.edu}

\affiliation{
Department of Electrical and Computer Engineering, University of Arizona, Tucson, Arizona 85721, USA
}
\affiliation{
J. C. Wyant College of Optical Sciences, University of Arizona, Tucson, Arizona 85721, USA}
\affiliation{
Ming Hsieh Department of Electrical and Computer Engineering, University of Southern California, Los
Angeles, California 90089, USA
}

\date{\today}

\begin{abstract}

The nature of dark matter is one of the most important open questions in modern physics. The search for dark matter is challenging since, besides gravitational interaction, it feebly interacts with ordinary matter. Mechanical sensors are one of the leading candidates for dark matter searches in the low frequency region. Here, we propose entanglement-enhanced optomechanical sensing systems to assist the search for DM with mechanical sensing devices. To assess the performance of our setup, we adopt the integrated sensitivity, which is particularly suitable for broadband sensing as it precisely quantifies the bandwidth-sensitivity tradeoff of the system. We then show that, by coherently operating the optomechanical sensor array and utilizing continuous-variable multi-partite entanglement between the optical fields, the array of sensors has a scaling advantage over independent sensors (i.e., $\sqrt{M}\rightarrow M$, where $M$ is the number of sensors) as well as a performance boost due to entanglement. Such an advantage is robust to imhomogeneities of the mechanical sensors and is achievable with off-the-shelf experimental components.

\end{abstract}

\maketitle

Identifying the nature of Dark Matter (DM) is one of the most pressing quests for fundamental physics research. The evidence for the existence and the particle nature of DM are ubiquitous---such as the cosmic microwave background survey~\cite{Springel:2006vs,aghanim2020planck}, gravitational lensing~\cite{Clowe:2006eq} and rotation curves of spiral galaxies~\cite{Rubin:1970zza,1975ApJ...201..327R,Rubin:1980zd,Bosma:1981zz}---with a consistent DM mass ranging over eighty orders of magnitude. 
Many searches focus on the particle-like regime, such as weakly interacting massive particles via direct detection searches. On the other hand, exploring the sub-eV regime for wave-like DM poses interesting experimental challenges and new opportunities for particle physics discovery. 

Depending on the DM model, various types of DM sensors have been designed. In axion DM model, DM can induce photons in a background magnetic field; therefore microwave cavities immersed in a powerful magnetic field are leveraged for a DM search~\cite{Sikivie:1983ip,ADMX:2019uok,ADMX:2021nhd,HAYSTAC:2018rwy,McAllister:2017lkb,CAPP:2020utb, girvin2016axdm,malnou2019,dixit2021,backes2021,Berlin2022hfx}.
In other models, DM induces forces on normal matter; therefore mechanical sensors good at sensing weak forces can be used for a DM search~\cite{dal2020resonators,dal2021VDM,carney2021DM,carney2021iop,monteiro2020PRL,moore2021levitated,afek2022trapped_sensors,Berlin2022hfx,Antypas2022asj,chowdhury2022membrane}. These new potential couplings could reveal how the relic density of DM comes about and possibly explain other puzzles of fundamental physics, such as the strong CP puzzle, the baryon-antibaryon asymmetry of our universe, and many others.

As the precision of sensors approaches the quantum regime, shot noise from vacuum fluctuations becomes relevant. To overcome vacuum noise, in the microwave cavity case, a squeezing enhanced DM search has been proposed~\cite{girvin2016axdm,malnou2019} and demonstrated~\cite{backes2021}.
Recently, arrays of microwave cavities~\cite{derevianko2018network,jeong2020prl,sikivie2020search,brady2022arxivDM} and optomechanical sensors~\cite{carney2021iop,carney2020PRD,windchime2022whitepaper} have been proposed to further enhance DM searches. The benefits of coherent post-processing with an optomechanical array operating at the standard quantum limit (SQL) have been discussed~\cite{carney2021iop,carney2020PRD,windchime2022whitepaper}, but explicit details---as well as potential advantages from entanglement-enhanced readout---remain largely unexplored. With backaction into play, it is unclear whether an optomechanical array can enjoy the entanglement advantages previously shown in the microwave case~\cite{brady2022arxivDM}.

\begin{figure}[t]
    \centering
    \includegraphics[width=\linewidth]{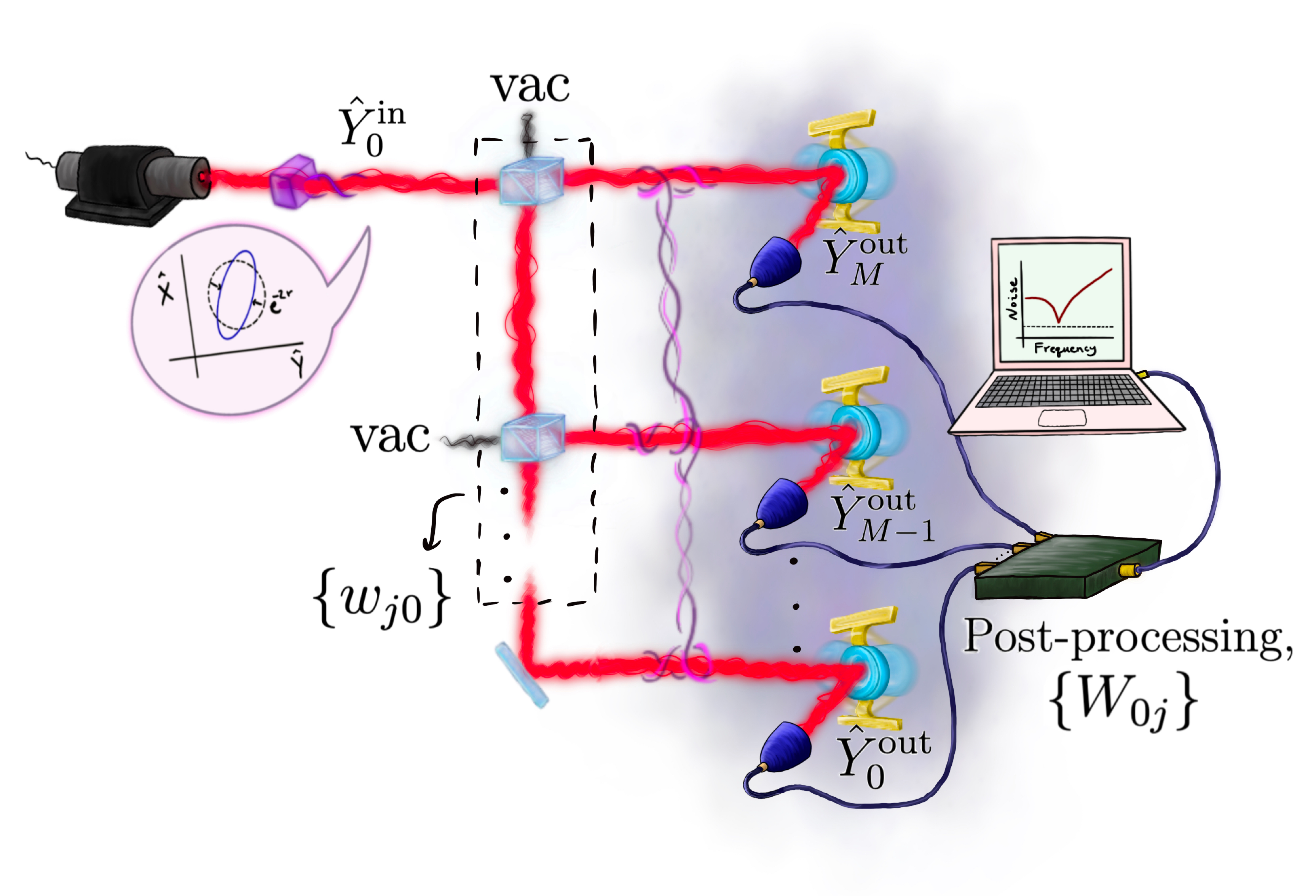}
    \caption{Illustration of our DQS proposal. A stochastic field (e.g., DM field; background cloud) faintly drives an array of mechanical oscillators, in a correlated manner, leading to feeble vibrations of the oscillators' positions. To interrogate the oscillators, squeezed laser light is distributed to the array via passive elements (beam-splitters, described by weights $\{w_{j0}\}$), which generates an entangled state. The impinging radiation reflects off the oscillators and is detected via homodyne detection. The measurement results are jointly combined via the weights $\{W_{j0}\}$, from which the signal is inferred.}
    \label{fig:dqs}
\end{figure}

In this work, we propose entanglement-enhanced readout of mechanical sensors for a DM search (see Fig.~\ref{fig:dqs}), exploiting recently developed techniques in distributed quantum sensing (DQS)~\cite{zhuang2018DQSCV,zhang2021dqs,xia2020demonstration,guo2020distributed}. Building upon the quantum theory of optomechanics~\cite{aspelmeyer2014RMP,bowen2015,barzanjeh2021OMrvw}, we show that, by coherently operating an array of mechanical sensors and utilizing continuous-variable multi-partite entanglement between the optical fields, entanglement enhancement and advantageous scaling with the number of sensors are simultaneously achievable. 

\QZ{
Our theory applies to a recent proof-of-principle experiment~\cite{xia2022} and futuristic optomechanical arrays, such as the proposed Windchime project for DM search~\cite{carney2020PRD,windchime2022whitepaper}, and may be relevant to other array-based proposals ~\cite{Afach:2018eze,Canuel:2020cxb}.
}

{\em Dark matter model.---}
The basic model of the driving force exerted on the mechanical oscillator (in the surface normal direction) by DM, with the mass density $\rho_{\rm DM}$, can be described by
$
F_{\rm dr}(t)\simeq F_{\rm DM} \cos(\Omega_{\rm DM} t+\varphi)
$
within a coherence time, with the amplitude of the drive force $F_{\rm DM}= g\sqrt{\rho_{\rm DM}} \calM$, where $g$ quantifies the coupling of DM to the mechanics for a particular DM model and $\calM$ characterizes material and geometrical properties of the sensor~\cite{Graham:2015ifn,carney2021DM,carney2021iop}. For the DM frequencies under consideration $\Omega_{\rm DM}\sim$ kHz (representing the dark matter particle's mass), the de Broglie wavelength of DM is around $10^5~$km; this large scale is pertinent for array-based setups that wish to take advantage of the signal correlations between the sensors within the array. Overall, the phase $\varphi$ is random and therefore $\hat{F}_{\rm dr}(t)$ is described by a stationary random process with a coherence time $1/\Delta_a$, and can be characterized by its power spectral density (PSD) $S_{F_{\rm dr}}\sim F_{\rm DM}^2/\Delta_a$. For any time-dependent operators $\hat{O}, \hat{O}'$ with stationary statistics, the PSD is defined as
\be 
{S}_{\hat{O} \hat{O}'}(\omega)=\frac{1}{2\pi}\int_{-\infty}^{+\infty}{\rm d}\omega' \ev{\hat{O}^{\dagger}(-\omega)\hat{O'}(\omega')} ,
\label{PSD_def}
\ee 
where $\hat{O}^{\dagger}(\omega)$ is the Fourier transform of the time-domain operator. We also define a symmetrized PSD, $\Bar{S}_{\hat{O} \hat{O}'}(\omega)\equiv[{S}_{\hat{O} \hat{O}'}(\omega)+{S}_{\hat{O} \hat{O}'}(-\omega)]/2$.
When $\hat{O} = \hat{O}'$, we simplify the notation as ${S}_{\hat{O}}(\omega)$.


{\em Simplified model.---} For clarity, we start our discussion with a simplified model for optomechanical sensors and then proceed to the full cavity model~\cite{aspelmeyer2014RMP,bowen2015,barzanjeh2021OMrvw}, which includes simplified model in the bad cavity limit. As shown in Fig.~\ref{fig:dqs}, each optomechanical sensor is composed of a 
mirror-like mechanical oscillator which couples dispersively to a free electromagnetic field. DM hypothetically couples to the mechanics and drives the oscillator's motion.


\QZ{
To detect the motion of the oscillator, one stimulates an input light field $\hat{E}^{\rm in}(t)$, which impinges on the mechanical element, and 
interferes with the output field $\hat{E}^{\rm out}(t)$ post-interaction. The mechanics completely reflects the input field and induces a small phase shift, $\zeta \hat{q}(t)\ll1$, such that $\hat{E}^{\rm in}(t)\to e^{{\rm i} \zeta \hat{q}(t)} \hat{E}^{\rm in}(t)$, where $\zeta=2\Omega_L/c$ and $\hat{q}$ is the position operator of the mechanics in meters. Setting the carrier frequency as zero, we can write out the strong laser mean field $E_0$ explicitly as $\hat{E}^{\rm in}(t)\approx E_0+\hat{E}(t)$; therefore, the output (including, e.g., detector efficiency, $0\leq\eta^2\le1$) can be expressed to leading order as
$
\hat{E}^{\rm out}(t)\approx\eta\Big[ E_0\big(1+{\rm i}\zeta \hat{q}(t)\big)+\hat{E}(t)\Big],
\label{Ein_out}
$
where loss-induced vacuum terms are omitted for simplicity. From here, we immediately see that the motion of the oscillator leads to a detectable optical displacement on the field $i\eta \zeta E_0 \hat{q}(t)$. Going to the Fourier domain, the phase quadrature of the filed has the input-output relation (see Appendix~\ref{app_single_sensor})
\be
\hat{Y}^{\rm out}(\omega)=\eta  \Big[\hat{Y}^{\rm in}(\omega)+\sqrt{2}E_0\zeta  \hat{q}(\omega)\Big],
\label{i_omega}
\ee
while the amplitude $\hat{Y}^{\rm out}(\omega)$ does not pick up the signal.
}

{\em Cavity optomechanics.---} 
We utilize linear input-output theory to describe radiation coupling to an optical cavity with one vibrating mirror (the mechanics). In this theory, the intra-cavity field and the mechanical motion of the mirror are dissipatively coupled to ingoing bath-modes---$(\hat{X}^{\rm in}, \hat{Y}^{\rm in})$ and $(\hat{Q}^{\rm in}, \hat{P}^{\rm in})$---at dissipation rates $\kappa$ and ${\gamma}$, respectively. The equations of motion for the open quantum system lead to a set of coupled first-order differential equations in the time domain for the intra-cavity modes ($\hat{X}$, $\hat{Y}$, $\hat{Q}$, and $\hat{P}$) in terms of the coupling rates and the ingoing bath modes, which can be analytically solved in the frequency domain. Here, $\hat{Q}=\hat{q}/\sqrt{2}q_{\rm zp}$ is the normalized position operator, where $q_{\rm zp}\equiv\sqrt{\hbar/2m\Omega}$ is the zero-point motion, and $\hat{P}$ is the conjugate momentum. The outgoing fluxes---denoted by the operators $(\hat{X}^{\rm out}, \hat{Y}^{\rm out})$ and $(\hat{Q}^{\rm out}, \hat{P}^{\rm out})$---are then determined via the input-output relations $\hat{X}^{\rm out}=\hat{X}^{\rm in}-\sqrt{\gamma}\hat{X}$ etc.; see Appendix~\ref{app:om_1sensor} and Refs.~\cite{aspelmeyer2014RMP,bowen2015} for further details.

An exact output relation for the spectral amplitude of the phase quadrature can be found~\cite{aspelmeyer2014RMP,bowen2015},
\begin{equation}
    \hat{Y}^{\rm out}(\omega) =-{\rm e}^{{\rm i}\varphi_\omega}\hat{Y}^{\rm in}(\omega) +2\sqrt{2{\gamma} C_\omega}\hat{Q}(\omega),\label{eq:y_out}
\end{equation}
where the phase $\varphi_\omega$ and the optomechanical cooperativity $C_\omega$ are defined via
\begin{align}
{\rm e}^{{\rm i}\varphi_\omega}\equiv \left(\frac{\kappa/2+{\rm i}\omega}{\kappa/2-{\rm i}\omega}\right), 
    C_\omega\equiv\frac{2G^2/
{\gamma}\kappa}{(1-2{\rm i}\omega/\kappa)^2}=\abs{C_\omega}{\rm e}^{{\rm i}\varphi_\omega}.\label{eq:cooperativity}
\end{align}
Here $G\equiv E G_0$ is the cavity-enhanced optomechanical coupling-rate, $G_0$ is the vacuum optomechanical coupling rate, and $E$ is the intra-cavity field (taken to be real). The intra-cavity field, $E$, is related to the amplitude of the input field, $E_0$, via $E^2=(4\kappa_r/\kappa^2)E_0^2$, where $\kappa$ is the total dissipation-rate of the cavity and $\kappa_{r}$ is the dissipation-rate to the readout port. Here, the spectral amplitude of the oscillator's position, $\hat{Q}(\omega)$, is given as,
\begin{equation}
    \hat{Q}(\omega) = 2\sqrt{{\gamma}}\chi_\omega \left(\hat{P}^{\rm in}(\omega)-\sqrt{2C_\omega}\hat{X}^{\rm in}(\omega)\right) + Q_{\rm dr}(\omega),\label{eq:q_intra}
\end{equation}
where $Q_{\rm dr}(\omega)$ is the diplacement induced by the driving force, $F_{\rm dr}$, and is related via $F_{\rm dr}(\omega)\equiv (\sqrt{\hbar m \Omega}/\chi_\omega) Q_{\rm dr}(\omega)$ and $\chi_\omega=\Omega/(\Omega^2-\omega^2-{\rm i}2\gamma\omega)$ is the mechanical susceptibility. The term proportional to the amplitude quadrature, $\hat{X}^{\rm in}$, represents the fluctuation of the oscillator's position due to radiation pressure.

With the complete cavity optomechanics model, we recover the mechanical motion induced quadrature displacement identified in Eq.~\eqref{i_omega} in the simplified model. The relation can be made quantitative  by treating the output mirror in the cavity model as a transparent window, see Appendix~\ref{app:om_1sensor} for further comparisons.

We estimate the force impressed on the mechanics from homodyne measurements on the phase-quadrature via
\begin{equation}
       \hat{F}(\omega)\equiv\frac{{\rm e}^{-{\rm i}\varphi_\omega/2}}{\chi_\omega}\sqrt{\frac{\hbar m\Omega}{8{\gamma}\abs{C_\omega}}}\hat{Y}^{\rm out}(\omega),\label{eq:force_estimator}
\end{equation}
which has units N/$\sqrt{\rm Hz}$. 
A general expression for the noise spectrum can also be derived,
\begin{multline}
    \bar{S}_{F_{\rm noise}}(\omega)= {\frac{\hbar m\Omega}{8{\gamma}\abs{C_\omega}\abs{\chi_\omega}^2}\bar{S}_{\hat{Y}^{\rm in}}(\omega)} +{8\hbar m{\gamma}\Omega\abs{C_\omega}\bar{S}_{\hat{X}^{\rm in}}(\omega)}
    \\+{\frac{2\hbar m\Omega}{\abs{\chi_\omega}}\Re\left(\frac{\chi_{\omega}}{\abs{\chi_\omega}}\tilde{S}_{\hat{X}^{\rm in} \hat{Y}^{\rm in}}(\omega)\right)} +{4\hbar m{\gamma}\Omega \bar{S}_{\hat{P}^{\rm in}}(\omega)},\label{eq:noise}
\end{multline}
where we have defined, 
\begin{equation}
    \tilde{S}_{\hat{X}^{\rm in}\hat{Y}^{\rm in}}(\omega)\equiv\frac{S_{\hat{X}^{\rm in}\hat{Y}^{\rm in}}(\omega)+S^{*}_{\hat{X}^{\rm in}\hat{Y}^{\rm in}}(-\omega)}{2}.\label{eq:s_tilde}
\end{equation}
The first term in Eq.~\eqref{eq:noise} is the shot noise, the second term is the back-action noise due to radiation pressure, the third term encodes the quadrature correlations, and the fourth term consists of mechanical fluctuations---e.g., $\bar{S}_{P^{\rm in}}\approx K_B T/\hbar\Omega$ for thermally dominated fluctuations. The SQL can be obtained by assuming initial vacuum fluctuations ($\bar{S}_{\hat{Y}^{\rm in}}=\bar{S}_{\hat{X}^{\rm in}}=1/2$ and $\tilde{S}_{\hat{X}^{\rm in}\hat{Y}^{\rm in}}=0$) and choosing ${\abs{C_\omega}=1/8\gamma\abs{\chi_\omega}}$, then (ignoring mechanical noise) the noise at the SQL is $\bar{S}_{F_{\rm noise}}^{\rm SQL}\equiv{\hbar m\Omega}/{\abs{\chi_\omega}}$. We can also incorporate detection loss $1-\eta^2$ in the cavity model via the simple substitution ${\bar{S}_{F_{\rm noise}}\rightarrow\bar{S}_{F_{\rm noise}}+\frac{1-\eta^2}{\eta^2}(\hbar m\Omega/16\gamma\abs{C_\omega}\abs{\chi_\omega}^2)}$; see Appendix~\ref{app:om_1sensor} for more discussion on loss. 

{\em Quantifying performance.---} 
Since the mass of the DM signal is \textit{a priori} unknown, one must integrate over many frequencies to rule out a range of potential masses for DM~\cite{Sikivie:1983ip,ADMX:2019uok,ADMX:2021nhd,HAYSTAC:2018rwy,McAllister:2017lkb,CAPP:2020utb,dixit2021,backes2021}. Hence, detection bandwidth of the setup is paramount, however sensitivity is equally important, as such is needed to quickly build statistical confidence in our measurements. A general figure of merit for broadband sensing of an incoherent force, which takes both sensitivity and bandwidth into account, is the integrated sensitivity,
\begin{equation}
    \mathcal{I}_{\Omega}\equiv\int_{0}^\infty\left(\frac{\bar{S}_{F_{\rm dr}}(\omega)}{\bar{S}_{F_{\rm noise}}(\omega)}\right)^2\frac{\dd\omega}{\pi}.\label{eq:sensitivity}
\end{equation}
For a thorough discussion about the integrated sensitivity being a good figure of merit in DM searches, see Refs.~\cite{Chaudhuri:2018rqn,chaudhuri2021optimal} and Appendix~\ref{sec:comments_DMsearch} for further discussion. Later, we evaluate the integrated sensitivity for an array of $M$ sensors and denote the quantity as $\mathcal{I}^{(M)}_\Omega$. 


Assuming $\bar{S}_{F_{\rm dr}}$ is approximately flat over the integration range (e.g., due to no prior information about DM mass~\footnote{With regards to a DM search, the hypothetical DM signal is sharply peaked around some frequency, the value of which is unknown. We assume no prior knowledge about where the DM signal may be in frequency space, and thus, each frequency bin is equiprobable to contain a signal. We can thus characterize the signal with a flat spectrum. The principal figure of merit is then the detector response over a large band of frequencies.}), we find the integrated sensitivity for the SQL, ${\mathcal{I}_{\Omega}^{\rm SQL}/\bar{S}^2_{F_{\rm dr}} ={4\gamma}/(\hbar m\Omega\gamma)^2}$, which is the ratio of the mechanical linewidth, $\gamma$, and the on-resonance PSD at the SQL, $\bar{S}^{\rm SQL}_{F_{\rm noise}}(\Omega)=\hbar m\Omega\gamma/2$. Without quantum resources, this sets the ultimate classical limit in broadband detection for a given set of mechanical parameters $m,\,{\gamma},\,\text{and}\,\Omega$, which in turn imposes limits on a DM search with a mechanical system.

{\em Optomechanical sensor array.---}
Squeezing the input radiation is known to increase the effective bandwidth in optomechanical sensing while leaving the peak sensitivity (set by the SQL) the same~\cite{chelkowski2005PRA,ulrik2018magnetometry,ma2017proposal,korobko2017bandsens,zhao2020prl_squeezing,mcculler2020prl_squeezing,yap2020epr_squeezing,sudbeck2020epr_squeezing,lough20216db}, thus resulting in squeezing-enhanced broadband sensing. Here, we extend the results on squeezing-enhanced broadband sensitivity to entanglement-enhanced broadband sensitivity with an array of $M$ mechanical sensors. In our setup, the optomechanics are not directly coupled across the array; rather, we allow for mixing of the input and output optical fields via linear optical elements (Fig.~\ref{fig:dqs}). We further suppose that the stochastic drive force (e.g., the DM field) impresses a correlated displacement on the sensors. We show that, by utilizing entangled optical fields to measure the mechanics, the squeezing-enhancement demonstrated for a single sensor~\cite{chelkowski2005PRA,ulrik2018magnetometry,ma2017proposal,korobko2017bandsens,zhao2020prl_squeezing,mcculler2020prl_squeezing,yap2020epr_squeezing,sudbeck2020epr_squeezing,lough20216db} naturally extends to a sensor-array with the same amount of squeezed photons. 

Consider $M$ input modes, $\{\hat{a}_{n}^{\rm in}\}_{n=0}^{M-1}$, and a strong laser field at frequency $\Omega_L$ on the $\hat{a}_0^{\rm in}$ mode, with non-trivial quantum fluctuations (e.g., squeezing) on the sidebands. This mode is mixed with the idling input-modes (consisting of uncorrelated vacuum fluctuations) of the remaining $M-1$ inputs via linear optical elements described by the dividing-weights $\{w_{n0}\}$, with $w_{n0}\in\mathbb{C}$. After interaction with the mechanics, we measure the outgoing phase quadrature at each sensor, $\hat{Y}_{n}^{\rm out}$, via homodyne detection (see Appendix~\ref{app:network} for more details). In post-processing, we convert the measurement result at the $n$th sensor to a force measurement via the relation~\eqref{eq:force_estimator} and write the resulting value as $\hat{F}_n$. We then statistically combine the signals from each sensor with combining-weights $\{W_{0n}\}$, with $W_{0n}\in\mathbb{C}$, and construct a weighted average force estimator,
\begin{equation}
     \hat{\overline{F}}(\omega)\equiv\sum_{n=0}^{M-1} W_{0n}\hat{F}_n(\omega).\label{eq:avg_estimate}
\end{equation}
In this manner, we capitalize on the correlations of the stochastic drive field (e.g., the spatial-uniformity of the DM field) across the array to achieve favorable scaling with the size of the array. 

Suppose that the drive force obeys the following statistics, {$\ev*{\hat{F}_{{\rm dr},  n}\hat{F}_{{\rm dr}, n^\prime}}=\mathcal{M}_n\mathcal{M}_{n^\prime}f^2$}, where $f$ is common to each sensor and $\mathcal{M}_n$ is a sensor dependent pre-factor (e.g., $f\sim \sqrt{\rho_{\rm DM}/\Delta_a}g$ for DM). The signal PSD of the force is then,
\begin{equation}
    \Bar{S}_{\hat{F}_{\rm dr}}^{(M)}(\omega)=\left|\sum_{n=0}^{M-1} W_{0n}\mathcal{M}_n\right|^2f^2.\label{eq:signal_weight}
\end{equation}
Consider the ideal scenario where all the sensors are identical. In this case, the dividing- and combining-weights are chosen to satisfy $\sum_k W^*_{0n}w_{k0}=\delta_{nk}$ and $\abs{w_{k0}}=\abs{W_{k0}}=1/\sqrt{M}$. In words, since the performance of each sensor is identical and the response of each sensor to the drive force is identical, the best strategy is to distribute the input radiation uniformly to each sensor  and then uniformly combine the signals.

\begin{figure}
    \centering
    \includegraphics[width=\linewidth]{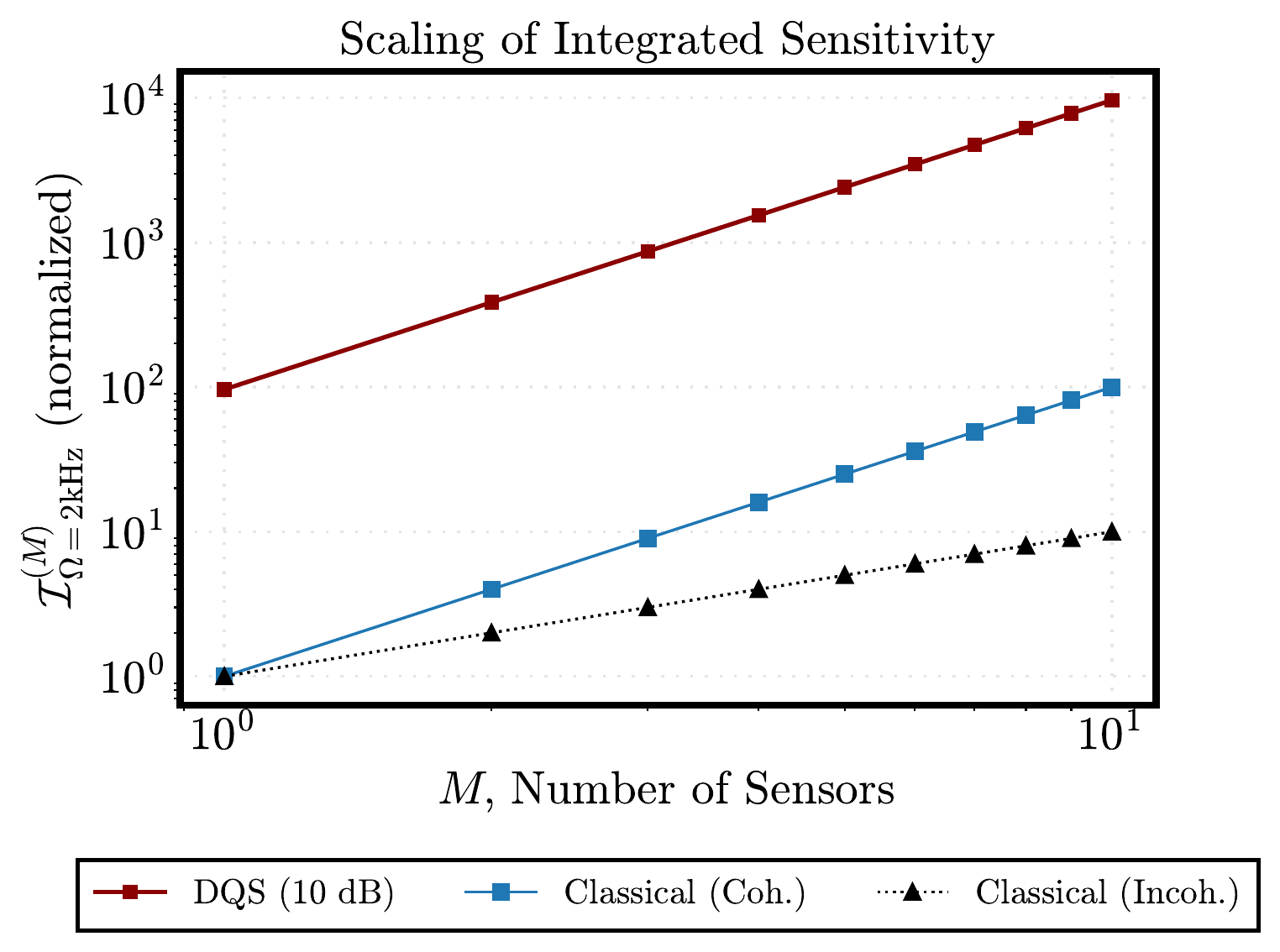}
    \caption{Scaling of the integrated sensitivity versus the number of mechanical sensors $M$. Red curve represents a DQS scheme with 10 dB input squeezing uniformly distributed across the array.  Blue curve represents a classical scheme with coherent post-processing (combining signals at the amplitude level). Black dotted curve represents a classical scheme that operates each sensor independently (combining signals incoherently at the power level).}
    \label{fig:M_scaling}
\end{figure}

The signal PSD in the ideal setting is $M$ times the single-sensor signal PSD ($\Bar{S}_{\hat{F}_{\rm dr}}^{(M)}=M\Bar{S}_{\hat{F}_{\rm dr}}$) which follows directly from Eq.~\eqref{eq:signal_weight}. This is due to the classical correlations of the drive field across the array. Moreover, if we scale the laser power with the number of sensors, $E_0^2\rightarrow ME_0^2$, such that the power per sensor is held fixed as we increase the number of sensors, then the multi-sensor force noise [Eq.~\eqref{eq:noise_network} of Appendix~\ref{app:network}] reduces to the single-sensor force noise of Eq.~\eqref{eq:noise}. Therefore, it follows that we can use an equal amount of squeezing in a multi-sensor setup as in a single-sensor setup to achieve an equivalent noise reduction; this comes along with an $M$-factor boost to the signal in the multi-sensor setup as previously mentioned.

\QZ{Our results are captured in Fig.~\ref{fig:M_scaling}, where we plot the integrated sensitivity for an array of identical sensors versus the number of sensors. To generate the data, we assume the mechanical sensors have similar parameters with Ref.~\cite{dal2021VDM} (see Appendix~\ref{app:exp})---resonance frequencies at $\Omega=2$ kHz, mechanical quality factors of $Q=10^9$, and masses of $m=6$ mg with each operating at $T=10$mK. The sensors act as an end mirrors for high-finesse optical cavities of $1$ mm lengths with $\kappa=.94$ GHz (corresponding to a finesse $\mathcal{F}\sim1000$). The input light has wavelength of $1.06\,\mu{\rm m}$, and the power \textit{per sensor} is chosen to be $P=2$ mW. 
}

For the DQS setup (red curve), a squeezed vacuum, with 10 dB squeezing ($N_s=2.03$ squeezed photons) and frequency dependent squeezing angle $\theta=\theta_\omega^\star$, is distributed uniformly across the array; see Refs.~\cite{zhao2020prl_squeezing,mcculler2020prl_squeezing,yap2020epr_squeezing,sudbeck2020epr_squeezing}, where tunability of the squeezing angle in optomechanical systems is addressed and Appendix~\ref{app:squeezing} for more information. We observe scaling enhancements for the DQS setup as well as for the classical setup with coherent processing (blue curve). The classical setup and the DQS setup enjoy a quadratic scaling enhancement over the independent sensor setup [Classical (Incoh.); black dotted] by leveraging spatial correlations of the drive field. On top of the scaling advantage, our DQS setup achieves a constant factor improvement over the classical scheme for all values of $M$, due to compounding the benefits of classical correlations from the drive field that boosts the signal and quantum correlations between the optical fields that results in a broadband reduction of the noise.

We stress that independent sensors cannot achieve the performance of our proposed DQS array with the same amount of squeezing---no matter the input laser power. Moreover, if we allow for joint post-processing but do not allow for entanglement between the modes, then $M$ independent squeezed vacua---each with $N_s$ number of squeezed photons---must be utilized (see Appendix~\ref{app:dqs_dcs}) in order to achieve the same performance as our DQS setup. This implies that the improvement in our DQS scheme is not necessarily due to the amount of squeezed light that impinges on a single mechanical oscillator but, rather, is a consequence of the quantum correlations between the optical fields that impacts the mechanics as a collective.


\QZ{
{\em Experimental projections.---} As an example to illustrate the benefit of entanglement-enhanced readout, Fig.~\ref{fig:dm_projection} provides projections (see Appendix~\ref{app:exp}) for the minimum detectable DM coupling strength $g_{\rm B-L}$ for a hypothetical DM search using the cm-scale silicon nitride membrane detector proposed in Ref.~\cite{dal2021VDM}. The subscript ${\rm B-L}$ indicates we have focused on ultralight DM coupled to `baryon minus lepton' charge, a typical choice of DM hypothesis for optomechanical systems~\cite{carney2021DM,carney2021iop,Antypas2022asj}. As shown in Fig.~\ref{fig:dm_projection}, we focus on a membrane mode resonating in the 1-10 kHz (1-100 peV) acoustic frequency (DM mass) range, where leading constraints come from MICROSCOPE~\cite{berge2018microscope,touboul2017microscope} (green shaded region), E\"{o}t-Wash~\cite{Wagner:2012ui} (gray shaded region), and LIGO/VIRGO~\cite{abbott2022constraints} (purple shaded region) experiments.
}

\QZ{Figure \ref{fig:dm_projection} assumes 20 cm square membrane detectors with the same parameters as in Fig. \ref{fig:M_scaling}, where a single sensor achieves a backaction-limited noise-equivalent acceleration resolution of $\sim 10^{-11} \text{ms}^{-2}\text{Hz}^{-1/2}$ on resonance. For an averaging time of 1 year, this corresponds to a minimum detectable DM coupling strength of $g_{\rm B-L}\sim 10^{-25}$ (light blue curve, $M=1$).
Classical sensor arrays improve the performance slightly, for both incoherent processing and coherent processing (darker blue curves, $M=10$). The entangled sensor-array generated from 10 dB squeezing is able to dramatically improve the performance (red solid), beating the SQL of classical sensor arrays (blue dotted). One can further increase and tune the power per sensor (to $P\approx10$ mW) to reach the DQS limit (red dotted) in the shot noise dominated region ($10^{-2}$-$10^{0}$ kHz) and further outperform existing constraints over a broad band.
}



\begin{figure}
    \centering
    \includegraphics[width=\linewidth]{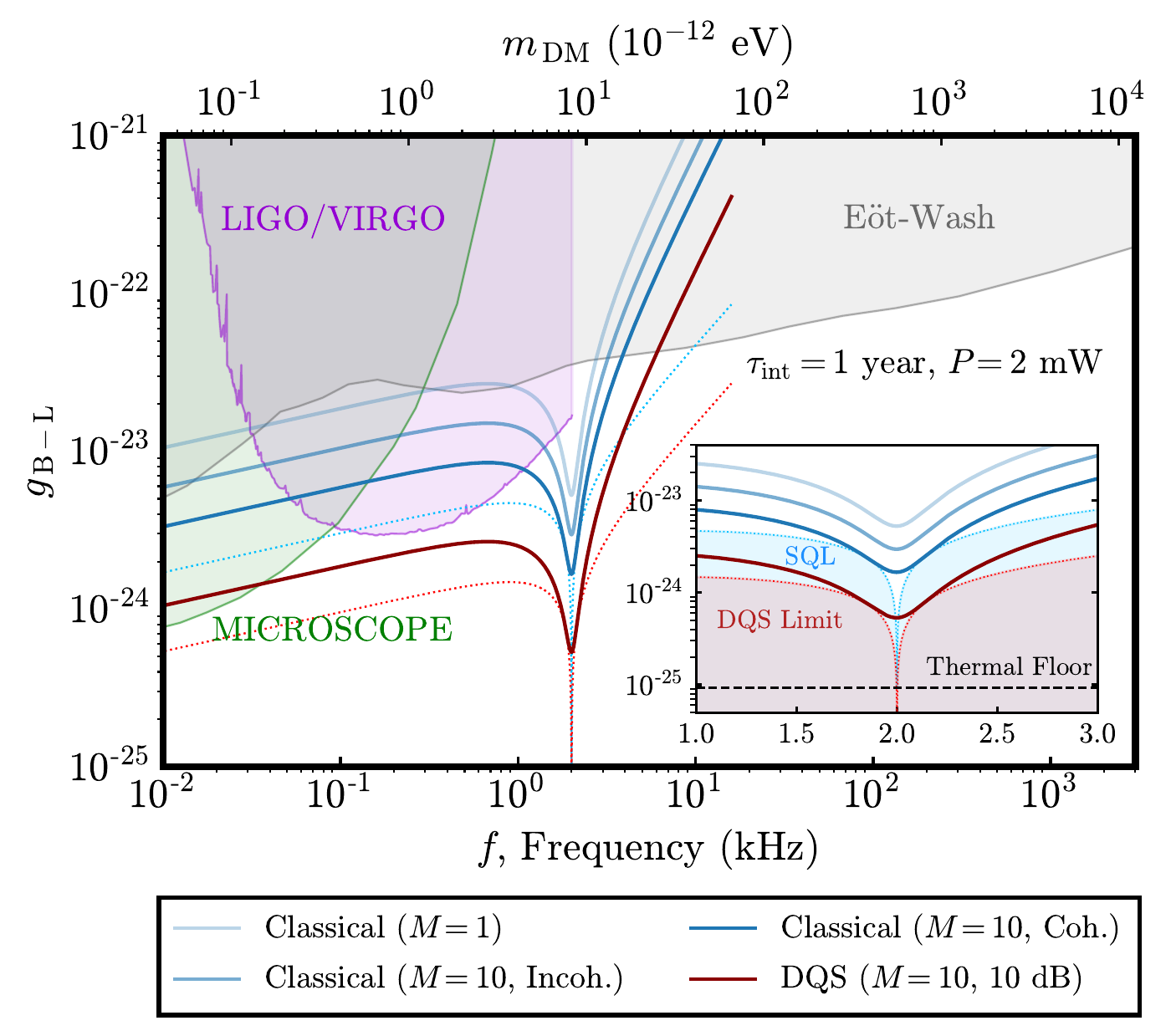}
    \caption{
    \QZ{Projected minimum detectable coupling strength of a hypothetical search for vector ultralight DM coupled to ${\rm B-L}$ charge~\cite{dal2021VDM} (see Appendix~\ref{app:exp}). The shaded regions indicate existing experimental constraints on coupling strength $g_{\rm B-L}$ from MICROSCOPE~\cite{berge2018microscope,touboul2017microscope}, E\"{o}t-Wash~\cite{Wagner:2012ui}, and LIGO/VIRGO~\cite{abbott2022constraints}. 
    }
    }
    \label{fig:dm_projection}
\end{figure}

\begin{acknowledgements}
This work is supported by NSF CAREER Award CCF-2142882, Defense Advanced Research Projects Agency (DARPA) under Young Faculty Award (YFA) Grant No. N660012014029, NSF Engineering Research Center for Quantum Networks Grant No. 1941583, NSF OIA-2134830 and NSF OIA-2040575, and U.S. Department of Energy, Office of Science, National Quantum Information Science Research Centers, Superconducting Quantum Materials and Systems Center (SQMS) under the contract No. DE-AC02-07CH11359. DJW, JM, and MC acknowledge support from the Northwestern University Center for Fundamental Physics and the John Templeton Foundation through a Fundamental Physics grant.

\end{acknowledgements}

\appendix

\section{Commutation relations and PSDs}
\label{appendix:definitions}
The commutation relation of the creation and annihilation operator of the optical field in the Heisenberg picture is $[\hat{a}^{\rm in}(t),\hat{a}^{{\rm in}\dagger}(t')]=\delta(t-t')
$. We define the quadratures of a mode as the real and imaginary parts of the annihilation operator $\hat{a}^{\rm in}$, respectively, as 
\begin{align}
&\hat{X}^{\rm in}(t)=[\hat{a}^{\rm in}(t)+\hat{a}^{{\rm in}\dagger}(t)]/\sqrt{2},
\\
&\hat{Y}^{\rm in}(t)=[\hat{a}^{\rm in}(t)-\hat{a}^{{\rm in}\dagger}(t)]/\sqrt{2}{\rm i}.
\end{align}
By taking Fourier transforms of the time domain fields, we define \begin{align}
\hat{X}^{\rm in}(\omega)=[\hat{a}^{\rm in}(\omega)+\hat{a}^{{\rm in}\dagger}(-\omega)]/\sqrt{2}, 
\\
\hat{Y}^{\rm in}(\omega)=[\hat{a}^{\rm in}(\omega)-\hat{a}^{{\rm in}\dagger}(-\omega)]/\sqrt{2}{\rm i},
\end{align}
where the annihilation operator $\hat{a}^{\rm in}(\omega)$ in the frequency domain is Fourier transform of $\hat{a}^{\rm in}(t)$ and obeys the commutation relation $[\hat{a}^{\rm in}(\omega),\hat{a}^{{\rm in}\dagger}(\omega')]=2\pi\delta(\omega-\omega')$.

In optomechanics, the PSD of the quadratures $\hat{X}^{\rm in}$, $\hat{Y}^{\rm in}$ are usually related to the back-action and shot noise. They can be calculated with Eq. (\ref{PSD_def}) of the main text. When the optical field is coherent, the PSD of the vacuum fluctuations are
\bal
S_{\hat{X}^{\rm in}}^{\rm vac}&=\int {\rm d}\omega' \frac{1}{2\pi}\left< \hat{X}^{{\rm in}\dagger}(-\omega)\hat{X}^{\rm in}(\omega') \right>\\&=\int {\rm d}\omega' \frac{1}{4\pi}\left< \hat{a}^{\rm in}(-\omega)\hat{a}^{{\rm in}\dagger}(\omega') \right>
\\&=\int {\rm d}\omega'\delta(\omega-\omega')=\frac{1}{2}
\\
S_{\hat{Y}^{\rm in}}^{\rm vac}&=\int {\rm d}\omega' \frac{1}{2\pi}\left< \hat{Y}^{{\rm in}\dagger}(-\omega)\hat{Y}^{\rm in}(\omega') \right>\\&=\int {\rm d}\omega' \frac{1}{4\pi}\left< \hat{a}^{\rm in}(-\omega)\hat{a}^{{\rm in}\dagger}(\omega') \right>
\\&=\int {\rm d}\omega'\delta(\omega-\omega')=\frac{1}{2},
\eal
When the optical field is in a two mode squeezed state (squeezed in the quadrature $\hat{X}^{{\rm in}\prime}$; see below), the annihilation operator $\hat{a}^{{\rm in}\prime}(\omega)$ is transformed by the
linear unitary Bogoliubov transformation
$\hat{a}^{{\rm in}\prime}(\omega)=\sqrt{N_S+1}\hat{a}^{\rm in}(\omega)+\sqrt{N_S}\hat{a}^{{\rm in}\dagger}(-\omega)$ and the quadrature operators $\hat{X}^{{\rm in}\prime}(\omega)$, $\hat{Y}^{{\rm in}\prime}(\omega)$ by the tranformations $\hat{X}^{{\rm in}\prime}(\omega)=(\sqrt{N_S+1}+\sqrt{N_S})\hat{X}^{\rm in}(\omega)$, $\hat{Y}^{{\rm in}\prime}(\omega)=\hat{Y}^{\rm in}(\omega)/(\sqrt{N_S+1}+\sqrt{N_S})$. The symmetrized PSD of the quadratures are
\bal
\bar{S}_{\hat{X}^{{\rm in}\prime}}^{\rm sqz}&=(\sqrt{N_S+1}+\sqrt{N_S})^2 \bar{S}_{\hat{X}^{\rm in}}\\&=\frac{(\sqrt{N_S+1}+\sqrt{N_S})^2}{2},
\\
\bar{S}_{\hat{Y}^{{\rm in}\prime}}^{\rm sqz}&=\frac{\bar{S}_{\hat{Y}^{\rm in}}}{(\sqrt{N_S+1}+\sqrt{N_S})^2} \\&=\frac{1}{2(\sqrt{N_S+1}+\sqrt{N_S})^2}.
\eal

\section{Aside remarks on DM search}
\label{app:dm_search}

\subsection{SNR and long observing runs}
In a DM search, primary objectives are to either measure the DM driving-force---therefore confirming the existence of DM---or (more likely) to exclude regions in physical parameter space (e.g., DM mass and coupling) where no such signal is found. This must be accomplished in the presence of various noise sources, whilst optimally utilizing resources (e.g., laser power, squeezing, etc.). 

For a single detection interval (of size $\sim1/\Delta_a$), the signal PSD of the DM-induced force is $\Bar{S}_{F_{\rm dr}}(\omega)\sim F_{ \rm DM}^2/\Delta_a$, where $F_{ \rm DM}$ is the amplitude of the (partially coherent) drive. For the optomechanical setup, the noise in the force estimation is given generally by Eq.~\eqref{eq:noise} of the main text in the main text. For simplicity, we approximate the Lorentzian PSD~\cite{dal2021VDM} of DM induced force as a delta-like peak of width $\Delta_a$, centered about the (unknown) mass of the DM particle.

Consider a long observation run of length $T_O$, such that $\Delta_aT_O\gg 1$, where $T_O$ is the total observation time, and decompose the total observation time into smaller intervals, each of size $\sim1/\Delta_a$. Statistically combining the results from each detection interval, we find the SNR over the entire observation,
\begin{equation}
    {\rm SNR} = \frac{\Bar{S}_{F_{\rm dr}}(\omega)}{\bar{S}_{F_{\rm noise}}(\omega)}\sqrt{\Delta_a T_O},\label{eq:SNR_TO}
\end{equation}
where the factor $\sqrt{\Delta_aT_O}$ is due to the law-of-large numbers. For longer observation times, $T_O$, smaller DM couplings can be probed at each frequency, $\omega$, thus providing a method to detect or exclude DM candidates.

\subsubsection{Random force perspective}
We now take a different viewpoint and consider long integration times, $T_{\rm int}\Delta_a\gg1$, such that the DM-induced force is incoherent and captured by the random force, $\xi_F(t)$, while the (partially coherent) deterministic drive $F_{\rm dr}(t)=0\,\forall\,t$. In this case, we take the ratio of the signal PSD (induced by the random force) and the other noise terms to obtain the SNR directly,
\begin{align}
{\rm SNR}
&=\frac{S_{\xi_F}(\omega) }{S_{\rm noise}(\omega)},
\label{eq:SNR_xi_F}
\end{align} 
As before, we consider a long observation time, such that $T_O\gg T_{\rm int}$, and then break the entire observation run into a large number of repetitions, $T_O/T_{\rm int}$. For fixed $T_O$, increasing $T_{\rm int}$ will not change the single-shot SNR in Eq.~\eqref{eq:SNR_xi_F} but will otherwise reduce the number of repetitions, resulting in a decreased SNR over the observation time, $T_O$. To increase the number of repetitions (and thus increase the SNR over the entire observing run), we could decrease the integration time, however we cannot do so arbitrarily, as the DM signal has a finite linewidth. Therefore, the best strategy is to set $T_{\rm int}\approx 1/\Delta_a$, where the precision is just enough to resolve DM signal. At this point, one can multiply the single-shot SNR of Eq.~\eqref{eq:SNR_xi_F} by the factor $\sqrt{T_O/T_{\rm int}}\approx\sqrt{T_O \Delta_a }$ from repeated measurements. Identifying $S_{\xi_F}(\omega)=S_{F_{\rm dr}}$, one then arrives at Eq.~\eqref{eq:SNR_TO}. Hence, from a data processing perspective, it is useful to think about the DM-induced force as a partially coherent drive, as discussed in the main text.

\subsection{Comments on figures of merit in a DM search}\label{sec:comments_DMsearch}
Through exhaustively thorough analyses, Refs.~\cite{Chaudhuri:2018rqn,chaudhuri2021optimal} established the integrated sensitivity as a primary figure of merit in a DM search. On the other hand, in original DM search proposals with microwave cavities~\cite{Sikivie:1983ip,ADMX:2019uok,ADMX:2021nhd,HAYSTAC:2018rwy,McAllister:2017lkb,CAPP:2020utb,dixit2021,backes2021}, a physically motivated figure of merit is the actual time that it takes to scan frequency space in search for a DM signal (by tuning the resonance frequency of the cavity)---which is the ``scan time'' for the DM search. 

In cavity setups, it can be shown that the scan rate (the inverse of the scan time) is proportional to the squared SNR, integrated over all resonance frequencies~\cite{kim2020revisit,malnou2019,brady2022arxivDM}. It turns out that, for a microwave cavity detector, the integrated sensitivity [defined in Eq.~\eqref{eq:sensitivity} of the main text] and the scan-rate are equivalent (up to a proportionality constant). This equivalence is due to the fact that the microwave cavity response (characterized by a Lorentzian profile) depends solely on the detuning from the resonance frequency of the cavity. This equivalence does not extend to optomechanical setups, due to the non-Lorentzian response of the mechanics. In our work, we follow the suggestions of Refs.~\cite{Chaudhuri:2018rqn,chaudhuri2021optimal} and thus take the integrated sensitivity as the preferred performance metric. The integrated sensitivity furthermore has an interpretation as the total amount of information about the signal over the entire frequency spectrum, which is important in broadband sensing scenarios, making this figure of merit precise and more generally useful; see, e.g., Ref.~\cite{polloreno2022broadband} for its use in a different context for two-level sensors.

The integrated sensitivity has also been considered in broadband detection of stochastic gravitational-wave backgrounds~\cite{thrane2013sensitivity,smith2019lisa,schmitz2021hep}. 

\section{Single sensor analyses of the simplified model}
\label{app_single_sensor}

\begin{figure}
    \centering
    \includegraphics[width=\linewidth]{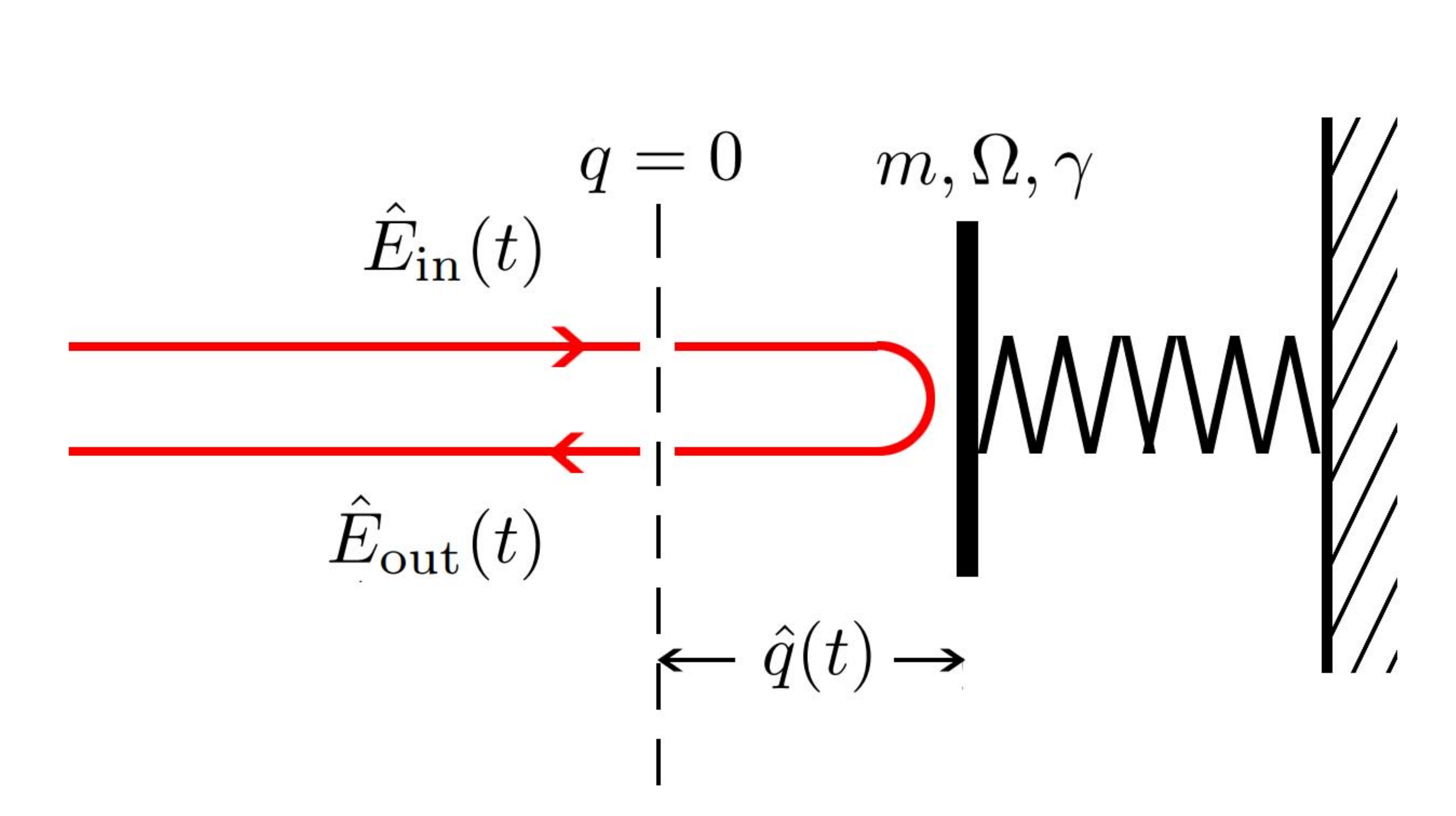}
    \caption{Schematic of the simplified model. $\hat{E}^{\rm in}$ and $\hat{E}^{\rm out}$ are the input and output field operators, respectively; $m$, $\Omega$, and $\gamma$ are  are the effective mass, resonance frequency, and damping rate of the mechanical oscillator, respectively; and $\hat{q}$ is the position operator of the mechanical oscillator. }
    \label{fig:2}
\end{figure}
In our simplified model, the optical field experience a phase shift caused by the motion of the mirror (as shown in Fig.~\ref{fig:2}). $\hat{E}^{\rm in}(t)$ and $\hat{E}^{\rm out}(t)$ are the input and output field operators at the equilibrium position ($q=0$) of the mirror, respectively. There is a phase rotation $2k\hat{q}(t')$ between the input and output field caused by the propagation. Here $k$ is the angular wavenumber of the field and $\hat{q}(t')$ is the position operator of the mirror at time $t'$. $t'$ is the time the wave arrives the mirror before its return back to position $q=0$ at time $t$. But because the motion speed of the mirror is much less than the light speed, we can treat $\hat{q}(t')\approx\hat{q}(t)$ approximately. The input-output relation of the field can be obtained as $\hat{E}^{\rm out}(t)= e^{2{\rm i}k \hat{q}(t)}\hat{E}^{\rm in}(t)=e^{{\rm i} \zeta \hat{q}(t)} \hat{E}^{\rm in}(t)$, where $\zeta=2\Omega_L/c$. With the input field decomposed in the frequency domain, considering the loss, the output field is 
\ba
\hat{E}^{\rm out}(t)&=& e^{{\rm i}\zeta\hat{q}(t)}\hat{E}^{\rm in}(t)\nonumber\\&\approx & e^{-{\rm i}\Omega_L t}\Big[\eta E_0\left(1+{\rm i}\zeta \hat{q}(t)\right)\nonumber\\&&\qquad+\int \frac{{\rm d}\omega}{2\pi} \eta \hat{a}^{\rm in}(\omega)e^{-{\rm i}\omega t}\Big] +\cdots,
\ea
The output field is measured via homodyne detection with the local oscillator field to be $\hat{E}^{\rm in}(t){\rm i}/E_0$. The detected quadrature is
\bal
\hat{Y}^{\rm out}(t)&=\hat{E}^{\rm out}(t)[-\hat{E}^{{\rm in}*}(t){\rm i}/\sqrt{2}E_0]+h.c.\\&=-\frac{{\rm i}}{\sqrt{2}}\Big[\eta E_0\left(1+{\rm i}\zeta \hat{q}(t)\right)\\&\qquad\qquad+\int \frac{{\rm d}\omega}{2\pi} \eta \hat{a}^{\rm in}(\omega)e^{-{\rm i}\omega t}+\cdots\Big]+h.c.\\&=-\frac{{\rm i}}{\sqrt{2}}\Big[{\rm i}2\eta E_0\zeta \hat{q}(t)+\int \frac{{\rm d}\omega}{2\pi} \eta \hat{a}^{\rm in}(\omega)e^{-{\rm i}\omega t}\\&\qqquad\qquad-\int \frac{{\rm d}\omega}{2\pi} \eta \hat{a}^{{\rm in}\dagger}(\omega)e^{{\rm i}\omega t}+\cdots\Big].
\eal
$\hat{Y}^{\rm out}$ in the frequency domain is obtained by Fourier transformation,
\bal
\hat{Y}^{\rm out}(\omega')&=\int \hat{Y}^{\rm out}(t)e^{{\rm i}\omega' t}{\rm d}t\\&=-\frac{{\rm i}}{\sqrt{2}} \eta [ \hat{a}^{\rm in}(\omega')-\hat{a}^{{\rm in}\dagger}(-\omega')]+\sqrt{2}\eta E_0\zeta \hat{q}(\omega')+\cdots\\&=\eta  \Big[\hat{Y}^{\rm in}(\omega')+\sqrt{2}E_0\zeta \hat{q}(\omega')\Big]+\cdots.
\label{Yout1}
\eal

The dynamics of the mechanical oscillator is determined by the radiation force, $\hat{F}_{\rm rad}$, from the light field probing the motion of the mechanics, the external driving force, $F_{\rm dr}$, exerted on the mechanics by the DM field and the random force, $\hat{\xi}_T$ induced by a thermal bath at temperature $T$. The motion equation of the mechanical oscillator is
\be 
m\ddot{\hat{q}}+2m\gamma\dot{\hat{q}}+k\hat{q}=F_{\rm dr}(t)+\hat{F}_{\rm rad}(t)+\sqrt{4 m \gamma K_B T}\hat{\xi}_T(t).
\ee
The oscillator position $\hat{q}$ can be obtained by solving the Langevin equation in the Fourier domain, resulting in the solution,
\begin{equation}
\hat{q}(\omega)=
\frac{\chi_\omega}{m\Omega}
\Big[F_{\rm dr}(\omega)+\hat{F}_{\rm rad}(\omega)+\sqrt{4 m \gamma K_B T}\hat{\xi}_T(\omega)\Big],
\label{xw}
\end{equation}
with $m$, $\gamma$, and $\Omega$ being the oscillator mass, mechanical damping rate, and mechanical resonance frequency, respectively. The random force $\hat{\xi}_T(\omega)$ has zero-mean and is Gaussian distributed, with $\langle \hat{\xi}_T(\omega)\hat{\xi}_T(\omega') \rangle=2\pi \delta(\omega-\omega')$, where $K_B$ is the Boltzmann constant. Finally, $\chi_\omega = \Omega/(\Omega^2-\omega^2-{\rm i}2\gamma\omega)$ is the complex linear response of the mechanics.

The radiation pressure force in time domain is given by
\ba
\hat{F}_{\rm rad}\left(t \right)&=&\kappa \hat{E}^{{\rm in}\,\dagger}(t)\hat{E}^{\rm in}(t)\nonumber\\&=&\kappa E_0\int \frac{{\rm d}\omega}{2\pi}  \hat{a}^{\rm in}(\omega)e^{-{\rm i}\omega t}+h.c.,
\label{rad}
\ea
where $\kappa=2\hbar\Omega_L/c$ is the momentum change of a photon reflected from the mirror. Its spectral amplitude is
\bal
\hat{F}_{\rm rad}\left(\omega' \right)&=\int \hat{F}_{\rm rad}\left(t \right)e^{{\rm i}\omega' t}{\rm d}t
=\kappa E_0 [\hat{a}^{\rm in}(\omega')+\hat{a}^{{\rm in}\dagger}(-\omega')]\\&=\sqrt{2}\kappa E_0\hat{X}^{\rm in}(\omega').
\label{radf}
\eal
Substituting Eq.~\eqref{xw} and Eq.~\eqref{radf} to Eq.~\eqref{Yout1}, we obtain
\begin{multline}
\hat{Y}^{\rm out}(\omega)=\eta  \Big\{\hat{Y}^{\rm in}(\omega)+\frac{\sqrt{2} E_0 \zeta \chi_\omega}{m\Omega}
\Big[\sqrt{2}\kappa E_0\hat{X}^{\rm in}(\omega)\\+F_{\rm dr}(\omega)+\sqrt{4 m \gamma K_B T}\hat{\xi}_T(\omega)\Big]\Big\}+\cdots.
\label{C8}
\end{multline}
The force estimator can be then expressed with the phase quadrature as
\ba
\hat{F}(\omega)&=&\frac{m\Omega}{\sqrt{2}\eta  E_0 \zeta \chi_\omega}\hat{Y}^{\rm out}(\omega)\nonumber\\&=&B(\omega)  \hat{Y}^{\rm in}(\omega)+\sqrt{2}\kappa E_0 \hat{X}^{\rm in}(\omega)+F_{\rm dr}(\omega)\nonumber\\&+&\sqrt{4 m \gamma K_B T}\hat{\xi}_T(\omega)\Big]+\cdots
\label{ia}
\ea
where
\be\label{eq:Bomega}
B(\omega)=\frac{m\Omega}{\sqrt{2} E_0\zeta \chi_\omega}.
\ee
from which the PSD of the force can be derived, 
\bal
S_{\hat{F}}(\omega)&=|B(\omega)|^2 \Big[S_{\hat{Y}^{\rm in}}(\omega)+\frac{1-\eta^2}{2\eta^2}\Big]+2\kappa^2 E_0^2 S_{\hat{X}^{\rm in}}(\omega)\\&+\frac{1}{2\pi}\int{\rm d}\omega' \ev{B^{*}(-\omega)\hat{Y}^{{\rm in}\dagger}(-\omega)\sqrt{2}\kappa E_0 \hat{X}^{\rm in}(\omega')}\\&+\frac{1}{2\pi}\int{\rm d}\omega' \ev{\sqrt{2}\kappa E_0 \hat{X}^{{\rm in}\dagger}(-\omega)B(\omega')\hat{Y}^{\rm in}(\omega')}\\&+4m\gamma K_B T+S_{F_{\rm dr}}(\omega)\\&=|B(\omega)|^2 \Big[S_{\hat{Y}^{\rm in}}(\omega)+\frac{1-\eta^2}{2\eta^2}\Big]+2\kappa^2 E_0 ^2 S_{\hat{X}^{\rm in}}(\omega)\\&+B'^{*}(-\omega)S_{\hat{Y}^{\rm in}\hat{X}^{\rm in}}(\omega)+B'(-\omega)S_{\hat{X}^{\rm in}\hat{Y}^{\rm in}}(\omega)\\&+4m\gamma K_B T+S_{F_{\rm dr}}(\omega),
\label{PSDiN}
\eal
where
\be
B'(\omega)=\sqrt{2}\kappa E_0 B(\omega).
\ee
In the derivation above, we have used the stationary statistic properties $\langle \hat{O}^{\dagger}(\omega)\hat{O}'(\omega')\rangle=f_{\hat{O} \hat{O}'}(\omega)\delta(\omega-\omega')$ for $\hat{O}$ and $\hat{O}'$ to be anyone of $\hat{X}^{\rm in}$ and $\hat{Y}^{\rm in}$. There is no correlations between the field quadrature, $F_{\rm dr}$ and $\hat{\xi}_T$, so the PSD of the force estimator is just the summation of the PSD of each of them. The PSD of force noise is just the summation of the residual terms apart from $S_{F_{\rm dr}}(\omega)$ on the right hand side of Eq.~\eqref{PSDiN}. Therefore the symmetrized PSD of noise is 
\begin{align}
 	&\bar{S}_{F_{\rm noise}}(\omega)	=4m\gamma K_B T+|B(\omega)|^2 \Big[\Bar{S}_{\hat{Y}^{\rm in}}(\omega)+\frac{1-\eta^2}{2\eta^2}\Big]\nonumber\\&+2\kappa^2 E_0 ^2 \Bar{S}_{\hat{X}^{\rm in}}(\omega)
 +\frac{B'^{*}(-\omega)S_{\hat{Y}^{\rm in}\hat{X}^{\rm in}}(\omega)}{2}	\nonumber
 	\\
 	&
 		+\frac{B'(-\omega)\left[S_{\hat{Y}^{\rm in}\hat{X}^{\rm in}}(-\omega)+S_{\hat{X}^{\rm in}\hat{Y}^{\rm in}}(\omega)\right]}{2}\nonumber
 	\\
 	& 		+\frac{B'(\omega)S_{\hat{X}^{\rm in}\hat{Y}^{\rm in}}(-\omega)}{2}\nonumber
\\&=4m\gamma K_B T+|B(\omega)|^2 \Big[\Bar{S}_{\hat{Y}^{\rm in}}(\omega)+\frac{1-\eta^2}{2\eta^2}\Big]\nonumber\\&+2\kappa^2 E_0 ^2 \Bar{S}_{\hat{X}^{\rm in}}(\omega)+2{\rm Re}[B'(-\omega)\tilde{S}_{\hat{X}^{\rm in}\hat{Y}^{\rm in}}(\omega)],
\label{psdnoise}
\end{align}
where we have used the properties  $B'(\omega)=B'^{*}(-\omega)$, $S_{\hat{O} \hat{O}'}(\omega)=S^{*}_{\hat{O}' \hat{O}}(\omega)$ in the derivation above.

\section{Single cavity optomechanics}\label{app:om_1sensor}

\subsection{Brief review of dynamics}
We start with the non-linear Hamiltonian of a cavity with one oscillating mirror. Let $\hat{X}$ and $\hat{Y}$ be the amplitude and phase quadratures, respectively, of the intra-cavity field and $\hat{Q}$ and $\hat{P}$ be the (dimensionless) position and momentum operators, respectively, of the movable mirror, such that $[\hat{X},\hat{Y}]=[\hat{Q},\hat{P}]={\rm i}$, with all other commutators vanishing. The Hamiltonian is then given by a sum of terms,
\begin{equation}
    \hat{H}=\hat{H}^{\rm free}+\hat{H}^{\rm int}+ \hat{H}^{\rm dr},
\end{equation}
where,
\begin{align}
    \hat{H}^{\rm free}&= \frac{\hbar\omega_c}{2}\left(\hat{X}^2+\hat{Y}^2\right)+\frac{\hbar\Omega}{2}\left(\hat{Q}^2+\hat{P}^2\right)\\
    \hat{H}^{\rm int}&=2\hbar G_0\hat{Q}\left(\hat{X}^2+\hat{Y}^2\right)\label{eq:int_ham}\\
    \hat{H}^{\rm dr}&= \hbar\mathcal{E}\hat{X}.
\end{align}
The first term is the free evolution of the intra-cavity quadrature operators (rotating at the cavity resonance-frequency, $\Omega_c=\Omega_L$, which we assume is on-resonance with the laser drive) plus the free-motion of the mirror (oscillating at the mechanical frequency, $\Omega$), the second term is the radiation-pressure interaction, which induces a shift of the cavity resonance-frequency depending on the mirror's position, and the last term is a linear-drive of the cavity field, which can be related to the input laser drive flux $E_0^2$ (see Ref.~\cite{bowen2015} for explicit expressions). Here, $G_0$ is the (normalized) vacuum optomechanical coupling rate. For a Fabry-Perot cavity of length $L$, $G_0=\frac{\Omega_c}{L}\sqrt{\frac{\hbar}{2m\Omega}}$, where $m$ is the mass of the oscillator.

One can linearise the Hamiltonian by expanding around the steady-state values of the intra-cavity field and the mirror's motion (the latter being induced by a strong field inside the cavity). We then go to the rotating frame of the cavity. This leads to the linear-approximation of the Hamiltonian in the rotating frame of the laser, 
\begin{equation}
    \hat{H}=\frac{\hbar\Omega}{2}\left(\hat{Q}^2+\hat{P}^2\right) + 2\hbar G\hat{Q}\hat{X},
\end{equation}
where $G\equiv E G_0$ is the cavity-enhanced optomechanical coupling-rate and $E$ is the intra-cavity amplitude (taken to be real). The intra-cavity amplitude is related to the input flux of the laser-drive, $E_0$, via,
\begin{equation}
    E^2=\frac{4\kappa_r}{\kappa^2}E_0^2,\label{eq:intra_in_power}
\end{equation}
where $\kappa$ is the total dissipation-rate of the cavity and $\kappa_{r}$ is the dissipation-rate to the readout port. If there is loss at a rate $\kappa_\ell$, then $\kappa=\kappa_r+\kappa_\ell$. Note that, given the laser field has frequency $\Omega_L$, the input laser-power is $P_{\rm in}=\hbar\Omega_LE_0^2\approx \hbar\Omega_L\kappa E^2/4$, where the approximation assumes over-coupling ($\kappa_r\approx\kappa$).

We include noise by assuming that the intra-cavity field and the mechanical motion of the mirror are dissipatively coupled to bath-modes---$(\hat{X}^{\rm in}, \hat{Y}^{\rm in})$ and $(\hat{Q}^{\rm in}, \hat{P}^{\rm in})$---at rates $\kappa$ and $\gamma$, respectively, such that $[\hat{X}_{\rm in}(t),\hat{Y}_{\rm in}(t^\prime)]=[\hat{Q}_{\rm in}(t),\hat{P}_{\rm in}(t^\prime)]={\rm i}\delta(t-t^\prime)$. In words, these modes represent an incoming photon-flux for the cavity field and an incoming phonon-flux for the oscillator's motion. The interactions with the baths lead to a coupled set of first-order, linear differential equations,
\begin{align}
    \dv{\hat{X}}{t}&=-\frac{\kappa}{2}\hat{X}+\sqrt{\kappa}\hat{X}_{\rm in},\label{eq:HL_1}\\
    \dv{\hat{Y}}{t}&=\frac{\kappa}{2}\hat{Y}+\sqrt{\kappa}\hat{Y}_{\rm in}-2G\hat{Q},\\
    \dv{\hat{Q}}{t}&=\Omega\hat{P},\\
    \dv{\hat{P}}{t}&=-\Omega\hat{Q}-2\gamma\hat{P}+2\sqrt{\gamma}\hat{P}_{\rm in}-2G\hat{X}. \label{eq:HL_4}
\end{align}
The outgoing fluxes can be found from the time-reversal of Eqs.~\eqref{eq:HL_1}-\eqref{eq:HL_4} and are related to the ingoing flux via the input-output relations $\hat{X}^{\rm out}=\hat{X}^{\rm in}-\sqrt{\kappa}\hat{X}$ etc., from which the (spectral) output amplitude and phase quadratures can be found,
\begin{align}
    \hat{X}^{\rm out}(\omega) &= -{\rm e}^{{\rm i}\varphi_\omega}\hat{X}^{\rm in}(\omega),\label{eq:x_out}\\
    \hat{Y}^{\rm out}(\omega) &=-{\rm e}^{{\rm i}\varphi_\omega}\hat{Y}^{\rm in}(\omega) +2\sqrt{2\gamma C_\omega}\hat{Q}(\omega),
\end{align}
with the latter expression agreeing with Eq.~\eqref{eq:y_out} of the main text and
\begin{equation}
    \hat{Q}(\omega) = 2\sqrt{{\gamma}}\chi_\omega \left(\hat{P}^{\rm in}(\omega)-\sqrt{2C_\omega}\hat{X}^{\rm in}(\omega)\right) + Q_{\rm dr}(\omega),
\end{equation}
where we have added an additional drive term $Q_{\rm dr}(\omega)$ due to a background, classical force.

\subsection{Bad cavity limit}
\label{appendix:bad_cavity_limit}
Here we show that, in the bad cavity limit, the simplified model (Appendix~\ref{app_single_sensor}) and the cavity optomechanics model agree. The goal is to show that the overall input-output relations agree. We accomplish this by comparing the force estimator of the simplified model, Eq.~\eqref{ia}, to the force estimator of the optomechanical model, Eq.~\eqref{eq:force_estimator} of the main text. Writing the latter out explicitly,
\begin{multline}
    \hat{F}(\omega) = \frac{{\rm e}^{-{\rm i}\varphi_\omega/2}}{\chi_\omega}\sqrt{\frac{\hbar m\Omega}{8{\gamma}\abs{C_\omega}}}\hat{Y}^{\rm out}(\omega) \\
    \hspace{1em}=-\frac{1}{\chi_\omega}\sqrt{\frac{\hbar m\Omega}{8{\gamma}\abs{C_\omega}}}\hat{Y}^{\rm in}(\omega) +\sqrt{4\hbar m\gamma\Omega}\hat{P}^{\rm in}(\omega) \\+F_{\rm dr}-\sqrt{8\hbar m\gamma\Omega C_\omega} \hat{X}^{\rm in}(\omega).
    \label{yout}
\end{multline}
In the bad cavity limit, ${\rm e}^{{\rm i}\varphi_\omega}\approx 1$ and $ C_\omega\approx 2G^2/
\gamma\kappa=8 E_0 ^2 G_0^2/\gamma\kappa^2$. If we shift the phase reference of the homodyne detection by $\pi$, we have $\hat{Y}^{\prime{\rm in}}(\omega)=-\hat{Y}^{\rm in}(\omega)$ and $\hat{X}^{\prime{\rm in}}(\omega)=-\hat{X}^{\rm in}(\omega)$, and Eq.~\eqref{yout} can be expressed as
\begin{multline}
    \hat{F}(\omega) =\frac{\sqrt{\hbar m \Omega}\kappa}{8 E_0  G_0\chi_\omega}\hat{Y}^{\prime{\rm in}}(\omega) +\sqrt{4\hbar m\gamma\Omega}\hat{P}^{\rm in}(\omega) \\+F_{\rm dr}+\frac{8 E_0 G_0}{\kappa}\sqrt{\hbar m\Omega} \hat{X}^{\prime{\rm in}}(\omega),\label{ia2}
\end{multline}
Comparing Eq.\eqref{ia} and Eq.\eqref{ia2} (at unity efficiency {$\eta =1$}), we make the following correspondence 
\be
\frac{\sqrt{\hbar m \Omega}\kappa}{8 E_0  G_0\chi_\omega}=B(\omega),
\ee
where $B(\omega)$ can be found in Eq.~\eqref{eq:Bomega}. This implies that
\be
\hbar\zeta=\frac{4G_0}{\kappa}\sqrt{2\hbar m\Omega}=\frac{4\hbar\Omega_L}{L\kappa}.
\ee
where we have used $E^2=(4/\kappa)E_0^2$ and taken $G_0
\sim\sqrt{\hbar/2m\Omega}\Omega_L/L$, which holds identically for a Fabry-Perot cavity. In the bad cavity limit of the cavity optomechanics model, the output mirror of the cavity is seen as a window (i.e., almost completely transparent) and thus $\kappa\sim c/L$, which is due to the fact that it roughly takes a time $L/c$ for a photon to leave the cavity after entry through the mirror.  Consequently, in this limit, the PSD expressions for each model agree.


\subsection{Squeezing enhancement}
\label{app:squeezing}

We introduce squeezing into the input optical field, which generally correlates the phase and amplitude quadratures and allows us to go below the SQL when interrogating the mechanics. Given a squeezing strength, $r$, and a squeezing angle, $\theta$, the PSDs of the input optical field are,
\begin{align}
    \bar{S}_{\hat{Y}^{\rm in}\hat{Y}^{\rm in}}(\omega)&=\frac{1}{2}\left({\rm e}^{-2r}\cos^2\theta+{\rm e}^{2r}\sin^2\theta\right),\label{eq:syy}\\
    \bar{S}_{\hat{X}^{\rm in}\hat{X}^{\rm in}}(\omega)&=\frac{1}{2}\left({\rm e}^{2r}\cos^2\theta+{\rm e}^{-2r}\sin^2\theta\right),\label{eq:sxx}\\
    \tilde{S}_{\hat{X}^{\rm in} \hat{Y}^{\rm in}}(\omega)&=\tilde{S}_{\hat{Y}^{\rm in} \hat{X}^{\rm in}}(\omega)=\frac{1}{2}\cos\theta\sin\theta({\rm e}^{2r}-{\rm e}^{-2r}).\label{eq:sxy}
\end{align}
We then assume the mechanical noise in the system is approximately flat, such that $\bar{S}_{P^{\rm in}P^{\rm in}}(\omega)=K_BT/\hbar\Omega$. Substituting these expressions into Eq.~\eqref{eq:noise} of the main text and rearranging, we obtain,
\begin{widetext}
\begin{equation}
    \bar{S}_{F_{\rm noise}}^{\rm sqz}(\omega)=\frac{\hbar m\Omega}{16{\gamma}\abs{C_\omega}\abs{\chi_\omega}^2}\left(\Big|\cos\theta-8{\gamma}\abs{C_\omega}\chi_\omega\sin\theta\Big|^2{\rm e}^{-2r}+\Big|\sin\theta+8{\gamma}\abs{C_\omega}\chi_\omega\cos\theta\Big|^2{\rm e}^{2r}\right)
    +4m{\gamma} K_B T,\label{eq:squeezed_noise}
\end{equation}
\end{widetext}
where we have used $\chi_\omega^*=\chi_{-\omega}$. In terms of the number of squeezed photons, $N_s$, we have ${\rm e}^{-2r}=1/(\sqrt{N_s}+\sqrt{N_s+1})^2$.

\begin{figure}[t]
    \centering
    \includegraphics[width=\linewidth]{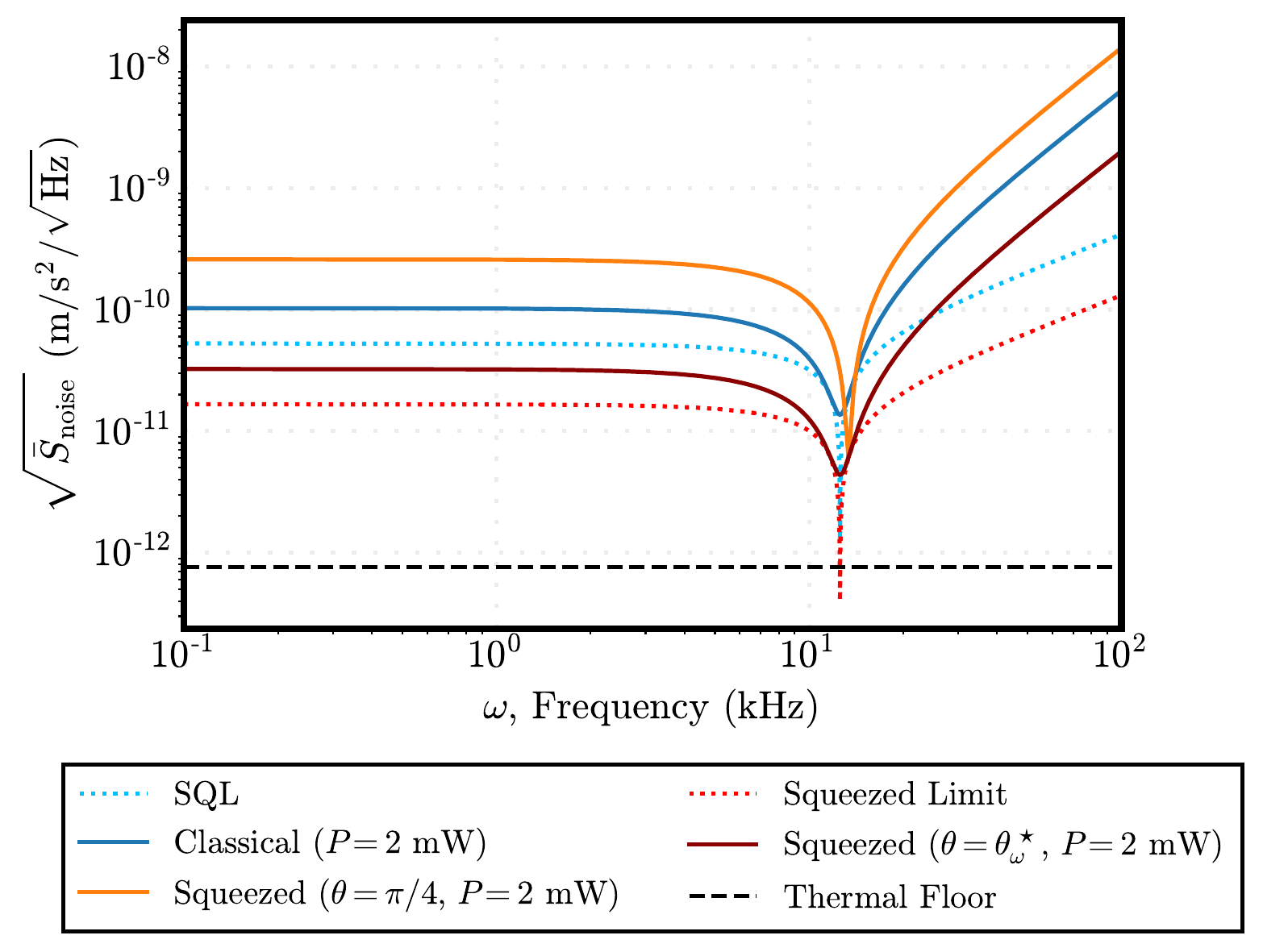}
    \caption{Acceleration noise (in units m/s$^2$/$\sqrt{\rm Hz}$) for a single optomechanical sensor. The amount of squeezing for each squeezed curve is taken as 10 dB (equivalent $N_s=2.03$ squeezed photons). Operational parameters are $\Omega=2\pi\times 2\,\rm{kHz}$, $T=10\,\rm{mK}$, $Q\equiv\Omega/2{\gamma}=10^9$, $\kappa=.94\,\rm{GHz}$, $m=6\rm{mg}$, and $G_0=46\,\rm{Hz}$ (cf.~\cite{dal2021VDM}). Angle $\theta_\omega^\star$ is chosen via $\theta_\omega^\star={\rm{argmax}}_\theta[\bar{S}_{\hat{F}_{\rm dr}}/\bar{S}_{F_{\rm noise}}]$.
    }
    \label{fig:noise_squeezed}
\end{figure}

For a fixed amount of squeezed photons and at a given frequency $\omega\neq\Omega$, there exists a laser-drive amplitude, $E^\star_\omega$, and a squeezing angle, $\theta^\star_\omega$, such that the force noise dips below the SQL. This can be seen by comparing the (dotted) red and blue curves in Fig.~\ref{fig:noise_squeezed}, which are the ultimate performances for classical and squeezing-enhanced detection (given a fixed squeezing at 10 dB), respectively.

In practice, the SQL and the squeezed-noise limit can only be reached at a particular frequency because the laser power, $E$, is constant (independent of $\omega$). Since we are concerned with broadband detection performance, we can otherwise choose the laser power to, e.g., optimize off-resonance detection; see the dark blue curve [Classical ($P=2$ mW)] in Fig.~\ref{fig:noise_squeezed}. 

Squeezing can beat classical limits in broadband detection if the squeezing angle, $\theta$, is appropriately chosen, a known result~\cite{chelkowski2005PRA,zhao2020prl_squeezing,mcculler2020prl_squeezing,ma2017proposal,yap2020epr_squeezing,sudbeck2020epr_squeezing}. This is especially pronounced if the squeezing angle is frequency tunable, such that we may operate the system at the optimal point, $\theta=\theta^\star_\omega$. We can observe this by inspecting the squeezed noises in Fig.~\ref{fig:noise_squeezed} [Squeezed limit, dotted red; Squeezed ($\theta=\theta^\star_\omega$, $P=2$ mW)] and comparing them to the classical detection schemes. Similar (though not as advantageous) benefits arise when the squeezing angle is not arbitrarily tunable but may nonetheless be chosen to take on a single value which maximizes the integrated sensitivity [Squeezed ($\theta=\pi/4,\,P=2$ mW), orange curve].

\begin{figure}
    \centering
    \includegraphics[width=\linewidth]{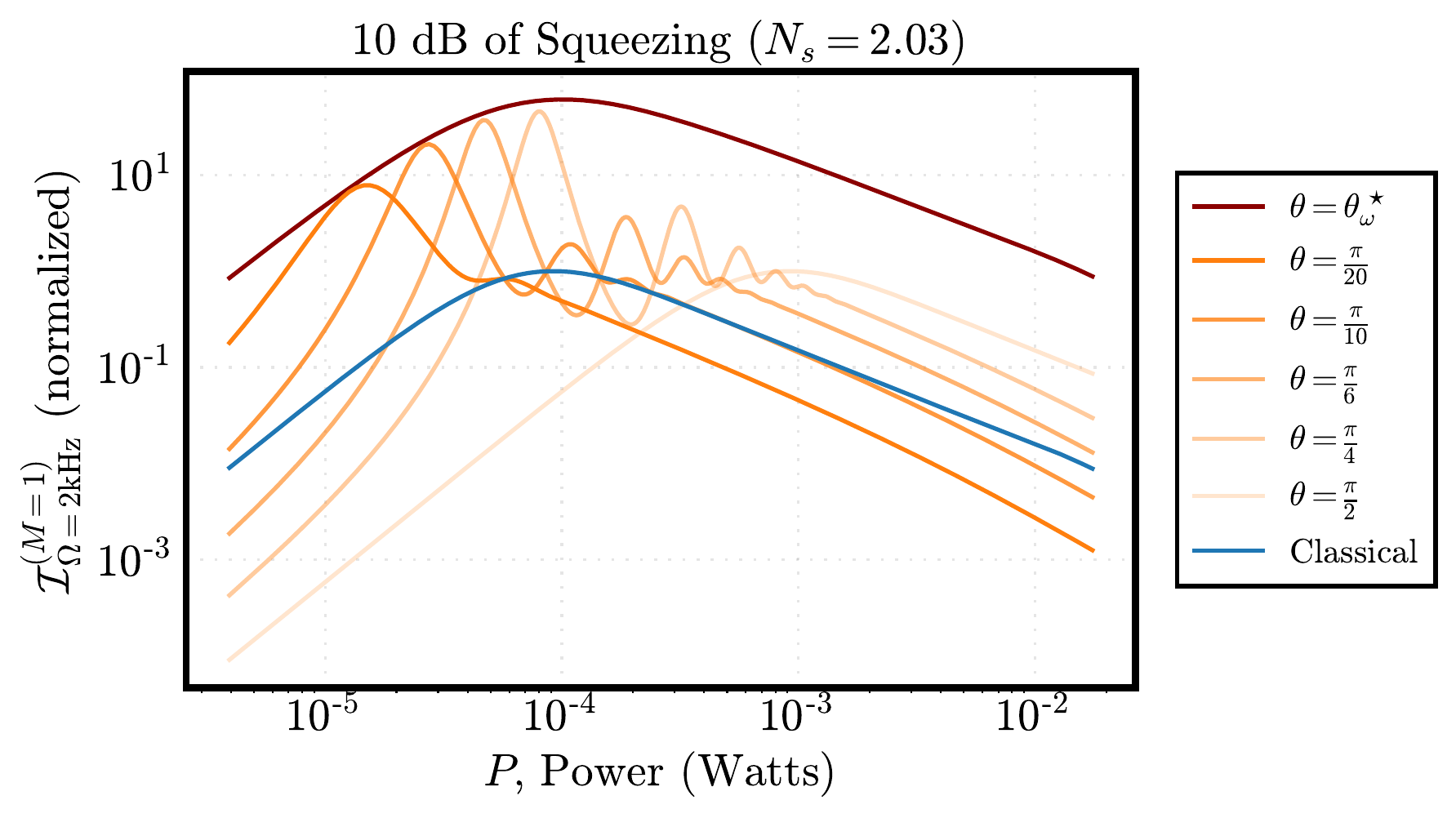}
    \caption{Single-sensor, integrated sensitivity for various detector configurations versus the input power, $P$. Squeezing level is 10 dB (equivalent $N_s=2.03$ squeezed photons) for each squeezed curve; $\theta$ is the squeezing angle. Operating parameters are the same as in Fig.~\ref{fig:noise_squeezed}. Here, $\theta_\omega^\star={\rm{argmax}}_\theta[\bar{S}_{\hat{F}_{\rm dr}}/\bar{S}_{F_{\rm noise}}]$.}
    \label{fig:int_sens}
\end{figure}

We now analyze broadband sensitivity for a single mechanical sensor, utilizing the integrated sensitivity as a performance metric. In Fig.~\ref{fig:int_sens}, we plot the integrated sensitivity, $\mathcal{I}_{\Omega}$, versus the input power for various detector configurations (classical and squeezed setups), assuming an approximately flat spectrum for the signal, $\bar{S}_{F_{\rm dr}}$. The squeezing angle $\theta_\omega^\star$ in Fig.~\ref{fig:int_sens} is chosen via $\theta_\omega^\star={\rm{argmax}}[\bar{S}_{\hat{F}_{\rm dr}}/\bar{S}_{F_{\rm noise}}]$; see Refs.~\cite{zhao2020prl_squeezing,mcculler2020prl_squeezing,yap2020epr_squeezing,sudbeck2020epr_squeezing}, where tunability of the squeezing angle in optomechanical systems is addressed. This optimal squeezing angle (which is frequency dependent) may not always be practically feasible. Hence, we also include fixed squeezing angles (orange curves). From the peaks in Fig.~\ref{fig:int_sens}, we see that, for a fixed power $P$, there exists an optimal fixed squeezing angle to work at that comes fairly close to the optimal sensitivity (red curve), at least for low powers.

\begin{figure}
    \centering
    \includegraphics[width=\linewidth]{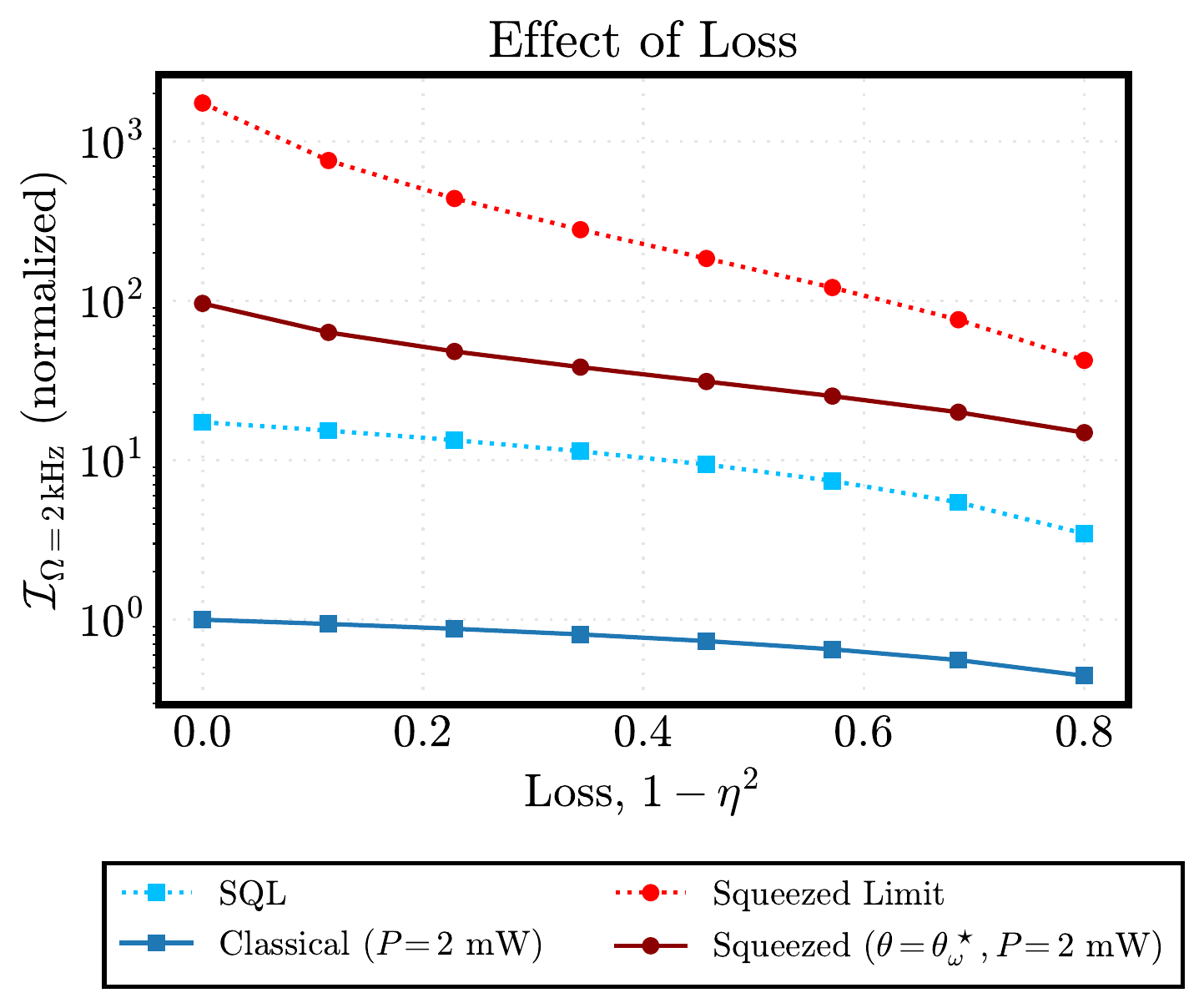}
    \caption{Integrated sensitivity versus loss, $1-\eta^2$.}
    \label{fig:loss_performance}
\end{figure}

Next, we consider the effect of loss on performance. For detection loss $1-\eta^2$, the force noise when loss is present can be found by the simple substitution, ${\bar{S}_{F_{\rm noise}}\rightarrow\bar{S}_{F_{\rm noise}}+\frac{1-\eta^2}{\eta^2}(\hbar m\Omega/16\gamma\abs{C_\omega}\abs{\chi_\omega}^2)}$. [We have thus factored out loss from the signal and pushed it in the noise, hence the $\eta^2$ in the denominator.] The first term is the force noise without loss, and the second term is additional measurement noise from the vacuum fluctuations of the loss port. As an example, for the SQL, $\bar{S}_{F_{\rm noise}}^{\rm SQL}\rightarrow\bar{S}_{F_{\rm noise}}^{\rm SQL}/\eta$; this can be shown by optimizing the force noise (assuming vacuum fluctuations for the input optical field), from which one finds the optimal cooperativity, ${\abs{C_\omega}=1/(8\eta\gamma\abs{\chi_\omega}})$, that leads to the aforementioned expression. In Fig.~\ref{fig:loss_performance}, we plot the integrated sensitivity for a mechanical oscillator with resonance frequency $\Omega=1\,\rm{kHz}$ versus loss, for various detection configurations (with and without squeezing). For $1-\eta^2>0$, the benefits from squeezing is ultimately limited by loss (as well as the thermal floor). Furthermore, the overall performance of each configuration degrades as the amount of loss increases, however squeezing is still beneficial for all non-zero values of the loss.

\section{Optmechanical array analysis}\label{app:network}
Below are detailed analyses of an array of optomechanical sensors, where we discuss some subtleties about the interaction Hamiltonian (describing the coupling between the mechanics and radiation) for an array, provide derivations of the force noise for an array of optomechanical sensors, discuss the SQL of the array, derive an explicit expression for the noise when input squeezed radiation is utilized, and, finally, derive an optimal squeezing angle for the array.

\subsection{Optomechanical interaction Hamiltonian for an array} \label{app:proof_int}
Consider the interaction Hamiltonian for the $j$th sensor [see, e.g., Eq.~\eqref{eq:int_ham}] in an optomechanical quantum array,
\begin{equation}
    \hat{H}^{\rm int}_j=4\hbar g_{0;j}\hat{Q}_j\left(\hat{a}_j^\dagger\hat{a}_j\right),\label{eq:int_hamj}
\end{equation}
where we have rewritten the expression in terms of the intra-cavity annihilation and creation operators, $\hat{a}_j$ and $\hat{a}_j^\dagger$. The annihilation operator in the Heisenberg picture is given in the Fourier domain by,
\begin{equation}
    \hat{a}_j=\int\dd\omega\hat{a}_j(\omega){\rm e}^{-{\rm i}\omega t}.\label{eq:a_j}
\end{equation}

We now linearise the interaction Hamiltonian by considering the intra-cavity field around its mean spectral amplitude, $\beta_{j}(\omega)$. This amplitude is linearly related to the input field amplitude, $\beta_{{\rm in};j}(\omega)$, via
\begin{equation}
    \beta_j(\omega)=\frac{\sqrt{\kappa_{r;j}}}{{\rm i}(\omega-\Omega_{c})+\kappa_j/2}\beta_{j}^{\rm in}(\omega),
\end{equation}
where $\Omega_{C}$ is the resonant frequency of the cavity, $\kappa_j$ is the total dissipation rate of the cavity, and $\kappa_{r;j}$ is the dissipation rate to the readout port. For simplicity, we assume all cavities in the array have approximately the same resonance frequency. Let $E_0(\omega)$ be the input laser-field amplitude to the primary mode (i.e., the $0$th input mode; all other modes are vacuum) prior to the beam-splitter array. Then, $\beta_{j}^{\rm in}(\omega)=w_{j0}(\omega)E_0(\omega)$ and
\begin{equation}
\begin{split}
    \beta_j(\omega)&=w_{j0}(\omega){\left(\frac{\sqrt{\kappa_{r;j}}}{{\rm i}(\omega-\Omega_c)+\gamma_j/2}E_0(\omega)\right)}\\
    &\equiv w_{j0}(\omega)E_j(\omega),
\end{split}
\end{equation}
where $E_j(\omega)$ is the intra-cavity amplitude if the entire laser field impinged on the cavity (i.e., in the absence of the power-dividing array). We now make the substitution $\hat{a}_j(\omega)\longrightarrow \hat{a}_j(\omega) + \beta_j(\omega)$
into Eq.~\eqref{eq:a_j}, using the equations above, and assuming the input field, $E_0(\omega)$, to be sharply peaked around the laser frequency, $\Omega_L$. We then find,
\begin{equation}
\begin{split}
    \hat{a}_j&\rightarrow\hat{a}_j+\int\dd\omega \,w_{j0}(\omega)E_j(\omega){\rm e}^{-{\rm i}\omega t}\\ &\approx \hat{a}_j+ w_{j0}(\Omega_L)E_j(\Omega_L){\rm e}^{-{\rm i}\Omega_C t}.
\end{split}
\end{equation}
We shall further assume that the laser-frequency is on-resonance with the cavities, such that $\Omega_L=\Omega_C$. Then, substituting the prior expression back into the interaction Hamiltonian of Eq.~\eqref{eq:int_hamj} and linearizing the result, we obtain, in the rotating reference frame of the cavity/laser,
\begin{equation}
\begin{split}
    \hat{H}^{\rm int}_j&\approx 4\hbar g_{0;j}\hat{Q}_j\abs{w_{j0}(\Omega_L)E_j}\left(\hat{a}_j{\rm e}^{-{\rm i}\Phi_j}+\hat{a}_j^\dagger{\rm e}^{{\rm i}\Phi_j}\right)\\
    &\equiv2\hbar g_{j}\hat{Q}_j\hat{X}_j(\Phi_j),
\end{split}
\end{equation}
where $\Phi_j\equiv \arg[E_0]+\arg[{w_{j0}(\Omega_L)}]$; $g_j\equiv g_{0;j}\abs{w_{j0}(\Omega_L)E_j}$; $\hat{X}_j(\Phi_j)\equiv\Re{\hat{a}_j{\rm e}^{-{\rm i}\Phi_j}}$; and we have discarded a term $\propto\hat{Q}_j\abs{w_{j0}(\Omega_L)E_j}^2$ which simply determines the steady-state of the mechanical oscillations. 

In gist, these results show that:
\begin{itemize}
    \item The bare coupling parameter of the $j$th sensor, $g_{0;j}$, gets enhanced to $g_j$, but only by a fraction of the total input laser field. Thus, if $C_j(\omega)$ denotes the mechanical cooperativity [as defined per Eq.~\eqref{eq:cooperativity} of the main text] of the $j$th sensor when the \textit{total} laser field, $E_0$, interacts with the mechanics, then the actual cooperativity of the mechanics is ${C_j^{\,\prime}}(\omega)=\abs{w_{j0}(\Omega_L)}^2C_j(\omega)$.
    \item From $\hat{X}_j(\Phi_j)$ and the definition of $\Phi_j$, we observe that the beam-splitter array can cause the quadrature bases of sensors within the array to differ, due to the angle $\arg[{w_{j0}(\Omega_L)}]$. This can cause problems when power-combining signals later on, as the signal at the $j$th sensor is encoded in the $j$th phase quadrature, $\hat{Y}_j(\Phi_j)$, and we want to align these quadratures in order to maximize signal output. To avoid potential mishaps here, we can choose $\arg[{w_{j0}(\Omega_L)}]=0\,\forall\,j$.
\end{itemize}

\subsection{Input-output relations}\label{app:in_out_network}
We now derive the input output relations for the sensor array depicted in Fig.~\ref{fig:dqs} of the main text, which leads to the general force noise in Eq.~\eqref{eq:noise_network} of the main text. A non-trivial input mode, $\hat{Y}_0^{\rm in}$, is mixed with vacuum fluctuations from $M-1$ idling input modes via a beam-splitter array governed by the weights $\{w_{k0}\}$, with $w_{k0}\in\mathbb{C}$. Just after the beam-splitter array, the phase-quadrature that impinges on the the $n$th mechanical-sensor is given as,
\begin{equation}
    \hat{Y}^{\prime~\rm in}_n= w_{n0}\hat{Y}_0^{\rm in} + \rm{vac},\label{eq:yprime_in}
\end{equation}
where ``vac'' indicates the (linear combination of) vacuum fluctuations from the $M-1$ idle input modes. After interaction with the mechanics, the phase quadrature at the output of the $n$th cavity is, which we measure via homodyne detection,
\begin{multline}
    \hat{Y}^{\rm out}_n=-{\rm e}^{{\rm i}\varphi_n(\omega)}\hat{Y}^{\prime~\rm in}_n +2\sqrt{2\gamma_{n} C^{\,\prime}_n(\omega)}\hat{Q}_n \\-8\gamma_{n} C^{\,\prime}_n(\omega) \chi_n(\omega) \hat{X}^{\prime~\rm in}_n,\label{eq:yprime_out}
\end{multline}
where $\hat{Q}_n$ is defined similarly as in Eq.~\eqref{eq:q_intra} of the main text, $\hat{X}^{\prime~\rm in}_n$ is defined likewise to Eq.~\eqref{eq:yprime_in} (which is required to preserve the canonical commutation relations between the phase and amplitude quadratures), and $C^{\,\prime}_n(\omega)\equiv C_n(\omega)\abs{w_{n0}(\Omega_L)}^2$, with $\Omega_L$ the laser frequency, which we have assumed to be the same for all sensors and assumed to be resonant with each cavity. Here, the cooperativity $C_n(\omega)$ [see Eq.~\eqref{eq:cooperativity} of the main text for an explicit expression] is defined with respect to the total power $E_0^2$. In our work, we take $\arg{[w_{n0}(\Omega_L)]}=0$ (or some constant independent of $n$). Without this choice, the quadrature bases of the output radiation at each sensor will not be aligned, which will lead to poorer performance when attempting to combine the signal amplitudes; see previous section for further details about this. 

After detection, the quadrature measurements can be converted to force measurements via the conversion formula~\eqref{eq:force_estimator} of the main text, from which the signal and noise PSDs can be derived.

\subsection{Noise Analysis}\label{sec:network_derivations}
From the previous section and Eq.~\eqref{eq:avg_estimate} of the main text, it follows that the PSD noise is,
\begin{equation}
    S_{\rm noise}(\omega)=\sum_{j,k}W^*_{0j}W_{0k}\ev{\hat{F}_j^\dagger\hat{F}_k}_{\rm noise},\label{eq:psd_estimate}
\end{equation}
where we have dropped frequency dependence for brevity. The subscript ``noise'' here indicates that we disregard the signal contribution to the PSD when evaluating this expression. The expression $\ev*{\hat{F}_j^\dagger\hat{F}_k}_{\rm noise}$ has four main terms: mechanical thermal noise, shot noise, back-action, and quadrature correlations. We consider the noise contributions in parts.

Mechanical thermal fluctuations at each sensor contribute to the total noise and obey the relations,
\begin{align}
        \ev{\hat{F}_j^\dagger\hat{F}_k}_{\Theta}&\equiv \frac{\sqrt{\hbar^2m_km_j\Omega_{k}\Omega_{j}}}{\chi_k\chi_j^*}\ev{\hat{Q}_k^\dagger\hat{Q}_j}_{\rm noise}\\
        &=4\sqrt{\hbar^2m_km_j\Omega_{k}\Omega_{j}\gamma_{k}\gamma_{j}}\underbrace{\ev{\hat{P}_j^{\rm in\,\dagger}\hat{P}_k^{\rm in}}}_{=\delta_{jk}(K_B T_k/\hbar\Omega_{k})}\\
        &=\delta_{jk}\left(4 m_k\kappa_kK_B T_k\right),
\end{align}
where $\Theta$ labels ``thermal" and Eq.~\eqref{eq:q_intra} of the main text was used to move from the first line to the second. The total thermal contribution to the PSD of Eq.~\eqref{eq:psd_estimate} is then, 
\begin{equation}
        S_{\rm noise}^\Theta\equiv\sum_{k=0}^{M-1}\abs{W_{0k}}^2\left(4 m_k\gamma_{k}K_B T_k\right),\label{eq:therm_network}
\end{equation}
which is just an average thermal noise over all sensors, according to the probability distribution $\abs{W_{0k}}^2$. [Recall, $0\leq\abs{W_{0k}}^2\leq1$ and $\sum_{k}\abs{W_{0k}}^2=1$.]

The fluctuations in the phase quadrature leads to shot-noise (SN), which is determined by,
\begin{equation}
        \ev{\hat{F}_j^\dagger\hat{F}_k}_{\rm SN}\equiv \frac{{\rm e}^{{\rm i}(\varphi_k - \varphi_j)/2}}{8\chi_k\chi_j^*}\sqrt{\frac{\hbar^2 m_km_j\Omega_{k}\Omega_{j}}{\gamma_{k}\gamma_{j}{\abs{C^{\,\prime}_k}\abs{C^{\,\prime}_j}}}}\ev{\hat{Y}^{\prime\,\rm in\dagger}_j\hat{Y}^{\prime\,\rm in}_k},\label{eq:F_SN_jk}
\end{equation}
We can expand the expectation value using Eq.~\eqref{eq:yprime_in} to write everything in terms of the PSDs of the input modes,
\begin{align}
        \ev{\hat{Y}^{\prime\,\rm in\dagger}_j\hat{Y}^{\prime\,\rm in}_k}&=\sum_{r,s}w_{jr}^*w_{ks}\underbrace{\ev{\hat{Y}^{\rm in\dagger}_r\hat{Y}^{\rm in}_s}}_{\propto\delta_{rs}}\\
        &=\sum_{r}w_{jr}^*w_{kr}\ev{\hat{Y}^{\rm in\dagger}_r\hat{Y}^{\rm in}_r}\\
        &=w_{j0}^*w_{k0} S_{Y_0^{\rm in}Y_0^{\rm in}}+{\left(\delta_{jk}-w_{j0}^*w_{k0}\right)S_{\rm vac}},\label{eq:sn_trouble}
\end{align}
where $S_{\rm vac}=1/2$ and we have used the unitary relation $\sum_{r=0}^{M-1}w_{jr}^*w_{kr}=\delta_{jk}$ (together with the assumption that the idle input modes all consist of vacuum fluctuations) to move from the second equality to the third equality. The first term in the last equality corresponds to the squeezed-vacuum on the primary (the $r=0$) mode, and the second term is a troublesome term which encodes the vacuum fluctuations from the other $M-1$ idling input modes. Defining the complex numbers, 
\begin{multline}
        \Delta_{jk}\equiv\\ {\rm e}^{{\rm i}(\varphi_k - \varphi_j)/2}\sqrt{\hbar^2 m_km_j\Omega_k\Omega_j}\left(\delta_{jk}-w_{j0}^*w_{k0}\right)W_{0j}^*W_{0k},
\end{multline}
and using $S_{\rm vac}=1/2$, we can write the force PSD due to shot noise generally as,
\begin{multline}
         S_{\rm noise}^{\rm SN}\equiv\abs{\left(\sum_{k=0}^{M-1}\frac{{\rm e}^{{\rm i}\varphi_k/2}}{2\chi_k}\sqrt{\frac{\hbar m_k\Omega_k}{\kappa_k{\abs{C^{\,\prime}_k}}}}W_{0k}w_{k0}\right)}^2S_{Y_0^{\rm in}Y_0^{\rm in}} \\ +\left(\sum_{j,k}\frac{\Delta_{jk}}{8\chi_k\chi_j^*}\sqrt{\frac{1}{\kappa_k\kappa_j{\abs{C^{\,\prime}_k}\abs{C^{\,\prime}_j}}}}\right).\label{eq:sf_sn}
\end{multline}
    
The fluctuations in the amplitude quadrature leads to back-action noise (BA), which is determined by,
\begin{multline}
        \ev{\hat{F}_j^\dagger\hat{F}_k}_{\rm BA}\equiv \\ 8{\rm e}^{{\rm i}(\varphi_k - \varphi_j)}\sqrt{\hbar^2 m_km_j\Omega_{k}\Omega_{j}\gamma_{k}\gamma_{j}\abs{C^{\,\prime}_k}\abs{C^{\,\prime}_j}}\ev{\hat{X}^{\prime\,\rm in\dagger}_j\hat{X}^{\prime\,\rm in}_k},
\end{multline}
with
\begin{equation}
        \ev{\hat{X}^{\prime\,\rm in\dagger}_j\hat{X}^{\prime\,\rm in}_k}
        =w_{j0}^*w_{k0} S_{X_0^{\rm in}X_0^{\rm in}}+{\left(\delta_{jk}-w_{j0}^*w_{k0}\right)S_{\rm vac}},\label{eq:ba_trouble}
\end{equation}
which follows from similar analyses that led to Eq.~\eqref{eq:sn_trouble}. We can then write the force noise due to back-action generally as,
\begin{multline}
        S_{\rm noise}^{\rm BA}\equiv \\ \abs{\left(\sum_{k=0}^{M-1}2{{\rm e}^{{\rm i}\varphi_k/2}}\sqrt{{2\hbar m_k\Omega_{k}\gamma_{k}\abs{C^{\,\prime}_k}}}W_{0k}w_{k0}\right)}^2S_{X_0^{\rm in}X_0^{\rm in}}\\ +\left(\sum_{j,k}4{\Delta_{jk}}\sqrt{{\gamma_{k}\gamma_{j}{\abs{C^{\,\prime}_k}\abs{C^{\,\prime}_j}}}}\right).\label{eq:sf_ba}
\end{multline}

One can likewise find an explicit expression for the quadrature correlation terms by using ${\ev{\hat{Y}^{\prime\,\rm in\dagger}_k\hat{X}^{\prime\,\rm in}_j}=w_{k0}^*w_{j0}\ev{\hat{Y}^{\rm in\dagger}_0\hat{X}^{\rm in}_0}}$. From which we can derive the contribution from quadrature correlations to the force PSD,
\begin{multline}
    S_{\rm noise}^{\rm corr}\equiv2\Re\Big[\left(\sum_{k=0}^{M-1}\frac{{\rm e}^{{-\rm i}\varphi_k/2}}{\chi_k^*}\sqrt{\frac{\hbar m_k\Omega_{k}}{\gamma_{k}{\abs{C^{\,\prime}_k}}}}W_{0k}^*w_{k0}^*\right)\\ \times\left(\sum_{j=0}^{M-1}{{\rm e}^{{\rm i}\varphi_j/2}}\sqrt{{\hbar m_j\Omega_{j}\gamma_{j}\abs{C^{\,\prime}_k}}}W_{0j}w_{j0}\right)\tilde{S}_{\hat{X}^{\rm in}_0 \hat{Y}^{\rm in}_0}\Big],\label{eq:corr_network}
\end{multline}
where $\tilde{S}_{\hat{X}^{\rm in}_0 \hat{Y}^{\rm in}_0}$ is defined in Eq.~\eqref{eq:s_tilde} of the main text. 

Finally, combining Eqs.~\eqref{eq:therm_network},~\eqref{eq:sf_sn},~\eqref{eq:sf_ba}, and~\eqref{eq:corr_network}, we derive a general expression of the force noise for the array,
\begin{widetext}
\begin{multline}
    \bar{S}_{F_{\rm noise}}^{(M)}=\\ \abs{\left(\sum_{k=0}^{M-1}\frac{{\rm e}^{{\rm i}\varphi_k/2}}{2\chi_k}\sqrt{\frac{\hbar m_k\Omega_{k}}{2\gamma_{k}{\abs{C^{\,\prime}_k}}}}W_{0k}w_{k0}\right)}^2\bar{S}_{Y_0^{\rm in}} + \abs{\left(\sum_{k=0}^{M-1}2{{\rm e}^{{\rm i}\varphi_k/2}}\sqrt{{2\hbar m_k\Omega_{k}\gamma_{k}\abs{C^{\,\prime}_k}}}W_{0k}w_{k0}\right)}^2\bar{S}_{X_0^{\rm in}} \\
     +2\Re\left[\left(\sum_{k=0}^{M-1}\frac{{\rm e}^{{-\rm i}\varphi_k/2}}{\chi_k^*}\sqrt{\frac{\hbar m_k\Omega_{k}}{\gamma_{k}{\abs{C^{\,\prime}_k}}}}W_{0k}^*w_{k0}^*\right)\left(\sum_{j=0}^{M-1}{{\rm e}^{{\rm i}\varphi_j/2}}\sqrt{{\hbar m_j\Omega_{j}\gamma_{j}\abs{C^{\,\prime}_k}}}W_{0j}w_{j0}\right)\tilde{S}_{\hat{X}^{\rm in}_0 \hat{Y}^{\rm in}_0}\right] \\+\sum_{k=0}^{M-1}\abs{W_{0k}}^2\left(4 m_k\gamma_{k}K_B T_k\right)+\bar{S}_{\rm res}^{(M-1)},
    \label{eq:noise_network}
\end{multline}
where
\begin{multline}
    \bar{S}_{\rm res}^{(M-1)}=\sum_{j,k}{\Delta_{jk}}\left(\frac{1}{8\chi_k\chi_j^*}\sqrt{\frac{1}{\kappa_k\kappa_j{\abs{C^{\,\prime}_k}\abs{C^{\,\prime}_j}}}} + 
    4\sqrt{{\gamma_{k}\gamma_{j}{\abs{C^{\,\prime}_k}\abs{C^{\,\prime}_j}}}}\right)
    \\= \sum_{k=0}^{M-1}\abs{W_{0k}}^2\left(\frac{\hbar m_k\Omega_{k}}{16\gamma_{k}{\abs{C^{\,\prime}_k}}\abs{\chi_k}^2}+4\hbar m_k\Omega_{k}\gamma_{k}\abs{C^{\,\prime}_k}\right)-\abs{\left(\sum_{k=0}^{M-1}\frac{{\rm e}^{{\rm i}\varphi_k/2}}{4\chi_k}\sqrt{\frac{\hbar m_k\Omega_{k}}{\gamma_{k}{\abs{C^{\,\prime}_k}}}}W_{0k}w_{k0}\right)}^2 \hspace{4.35em}\\
    -\abs{\left(\sum_{k=0}^{M-1}2{{\rm e}^{{\rm i}\varphi_k/2}}\sqrt{{\hbar m_k\Omega_{k}\gamma_{k}\abs{C^{\,\prime}_k}}}W_{0k}w_{k0}\right)}^2.\label{eq:residual_noise}
\end{multline}
\end{widetext}
In Eq.~\eqref{eq:noise_network}, the first three terms are due to the shot noise, back-action noise, and quadrature correlations, respectively, of the non-trivial input mode, $\hat{a}_0^{\rm in}$; these generalize the single-sensor, quadrature noises of Eq.~\eqref{eq:noise} of the main text to an array of sensors. The fourth term is the weighted average of the independent mechanical fluctuations of the various sensors. The final term [written out explicitly in Eq.~\eqref{eq:residual_noise}] contains the residual vacuum fluctuations from the $M-1$ idling input modes. These residual vacuum fluctuations do not contribute much to the noise, so long as the resonance frequencies of the mechanical systems are nearly identical, $\Omega_k\approx\Omega\,\forall\,k$; see below for more discussion on this. 

From Eq.~\eqref{eq:noise_network}, we can assess the performance of the array for any input radiation to the $\hat{a}_0^{\rm in}$ mode. For instance, the SQL for the array is readily obtained by setting the input to vacuum noise. Doing so, it is straightforward to show that the SQL noise for the array is a weighted average (with respect to the distribution, $\abs{W_{0k}}^2$) of the SQL noises for the individual sensors. Similarly, when squeezing is present in the input field, an expression for the noise in terms of the squeezing parameters can be found, from which we can, e.g., derive a formula for the squeezing angle that approximately cancels anti-squeezing noise; see the next section for details on squeezing. Finally, for a set of identical sensors, it is easy to show that Eq.~\eqref{eq:noise_network} reduces to the single-sensor noise of Eq.~\eqref{eq:noise} in the main text (i.e., the residual vacuum fluctuations vanish identically), upon taking $w_{k0}=W_{0k}^*$ and $\abs{W_{0k}}=1/\sqrt{M}$.

\paragraph{Residual vacuum fluctuations}
We make a few comments about power distribution and the residual vacuum fluctuations from idling input modes of the array [i.e., $\bar{S}_{\rm res}^{(M-1)}$ of Eq.~\eqref{eq:residual_noise}]. We can choose the distribution weights, $\{w_{k0}\}$, to minimize the residual vacuum fluctuations, but in general, we can not eradicate the vacuum noise entirely. The reason being that, in Eq.~\eqref{eq:residual_noise}, there is a phase difference between the third term and the fourth term---specifically due to the phase of the complex mechanical susceptibility, $\arg(\chi_k)$. This generally restricts us from eliminating the residual shot noise and back-action noise simultaneously [first and second terms in Eq.~\eqref{eq:residual_noise}, respectively], however we can get rid of one \textit{or} the other by choosing $w_{k0}$ appropriately, ultimately leaving some small amount of residual noise which depends on the phases $\arg(\chi_k)$. On the one hand, this is not much of an impediment near resonance, since $\arg{\chi_k(\Omega)}=\pi/2\,\forall\,k$ nor is it an issue far off resonance, where the susceptibility is approximately real. Thus, in these regimes, the residual vacuum noises can be approximately canceled. This, of course, assumes that the resonance frequencies of the mechanical systems are almost identical, $\Omega_k\approx\Omega\,\forall\,k$.

\subsection{Squeezing the array}
Assuming that the primary input mode to the array is squeezed (see Fig.~\ref{fig:dqs} of the main text), such that Eqs.~\eqref{eq:syy}-\eqref{eq:sxx} are satisfied, the force noise of Eq.~\eqref{eq:noise_network} can be written in terms of the squeezing strength, $r$, and squeezing angle, $\theta$,
\begin{widetext}
\begin{multline}
    \bar{S}_{F_{\rm noise}}^{{\rm sqz}, (M)}= \frac{1}{2}\abs{\left(\sum_{k=0}^{M-1}\frac{{\rm e}^{{\rm i}\varphi_k/2}}{2\chi_k}\sqrt{\frac{\hbar m_k\Omega_{k}}{2\gamma_{k}{\abs{C^{\,\prime}_k}}}}W_{0k}w_{k0}\right)\cos\theta-\left(\sum_{k=0}^{M-1}2{{\rm e}^{{\rm i}\varphi_k/2}}\sqrt{{2\hbar m_k\Omega_{k}\gamma_{k}\abs{C^{\,\prime}_k}}}W_{0k}w_{k0}\right)\sin\theta}^2{\rm e}^{-2r}\\
    + \frac{1}{2}\abs{\left(\sum_{k=0}^{M-1}\frac{{\rm e}^{{\rm i}\varphi_k/2}}{2\chi_k}\sqrt{\frac{\hbar m_k\Omega_{k}}{2\gamma_{k}{\abs{C^{\,\prime}_k}}}}W_{0k}w_{k0}\right)\sin\theta+\left(\sum_{k=0}^{M-1}2{{\rm e}^{{\rm i}\varphi_k/2}}\sqrt{{2\hbar m_k\Omega_{k}\gamma_{k}\abs{C^{\,\prime}_k}}}W_{0k}w_{k0}\right)\cos\theta}^2{\rm e}^{2r}\\
    +\sum_{k=0}^{M-1}\abs{W_{0k}}^2\left(4\hbar m_k\gamma_{k}K_B T_k\right)
    +\bar{S}_{\rm res}^{(M-1)},\label{eq:sq_noise_network}
\end{multline}
\end{widetext}
where a specific form of the residual vacuum noises, $\bar{S}_{\rm res}^{(M-1)}$, is in Eq.~\eqref{eq:residual_noise}. We can choose the squeezing angle, $\theta$, to approximately cancel the anti-squeezed noise (the term proportional to ${\rm e}^{2r}$),
\begin{equation}
    \tan\theta_\omega^\star=-\frac{\abs{\sum_{k=0}^{M-1}8{\rm e}^{{\rm i}\varphi_k/2}\sqrt{{\hbar m_k\Omega_{k}\gamma_{k}\abs{C^{\,\prime}_k}}}W_{0k}w_{k0}}}{\abs{\sum_{k=0}^{M-1}{\rm e}^{{\rm i}\varphi_k/2}\frac{\Omega_k^2-\omega^2}{\Omega_k}\sqrt{\frac{\hbar m_k\Omega_{k}}{\gamma_{k}{\abs{C^{\,\prime}_k}}}}W_{0k}w_{k0}}},
\end{equation}
where we have made explicit that this optimal squeezing angle is frequency dependent. For identical sensors (or a single sensor), this expression reduces to that found in standard texts; see, e.g., Section 5.4.2 in Ref.~\cite{bowen2015}.

\subsection{Classical limits and SQL}\label{sec:sql_network}
If we assume initial vacuum fluctuations in all input quadratures, then the total noise at the output is simply an average of each individual sensor noise, weighted by the distribution $\abs{W_{0k}}^2$. Concretely,
\begin{multline}
    \bar{S}_{F_{\rm noise}}^{\,\rm cl}(\omega)= \sum_{k=0}^{M-1}\abs{W_{0k}}^2\left(4\hbar m_k\gamma_{k}K_B T_k\right) \\ + \sum_{k=0}^{M-1}\abs{W_{0k}}^2\left(\frac{\hbar m_k\Omega_{k}}{16\gamma_{k}{\abs{C^{\,\prime}_k}}\abs{\chi_k}^2} +4\hbar m_k\Omega_{k}\gamma_{k}\abs{C^{\,\prime}_k}\right) \\
\end{multline}
where ``cl'' stands for classical. The SQL for the array is then the average SQL of the sensors. This is found by setting $\abs*{C^{\,\prime}_k}=1/8\gamma_{k}\abs{\chi_k}$ in the above and summing the shot noise and back-action, resulting in (ignoring the mechanical fluctuations momentarily),
\begin{equation}
    \bar{S}_{F_{\rm noise}}^{\rm SQL}(\omega)\equiv\sum_{k=0}^{M-1}\abs{W_{0k}}^2\left(\frac{\hbar m_k\Omega_{k}}{\abs{\chi_k}}\right).
\end{equation}
The total noise is then the noise at the SQL plus the thermal fluctuations of the mechanics.

We note that operating at the each sensor independently at the SQL but allowing for joint post-processing is a distributed classical sensing (DCS) scheme which allows for better scaling with the size of the array than independent sensors. Without squeezing/entanglement, the peak performance of a purely classical setup is set by such a CL-DCS scheme.  

\subsection{DQS vs. DCS}\label{app:dqs_dcs}

\begin{figure*}[t]
    \centering
    \includegraphics[width=\linewidth]{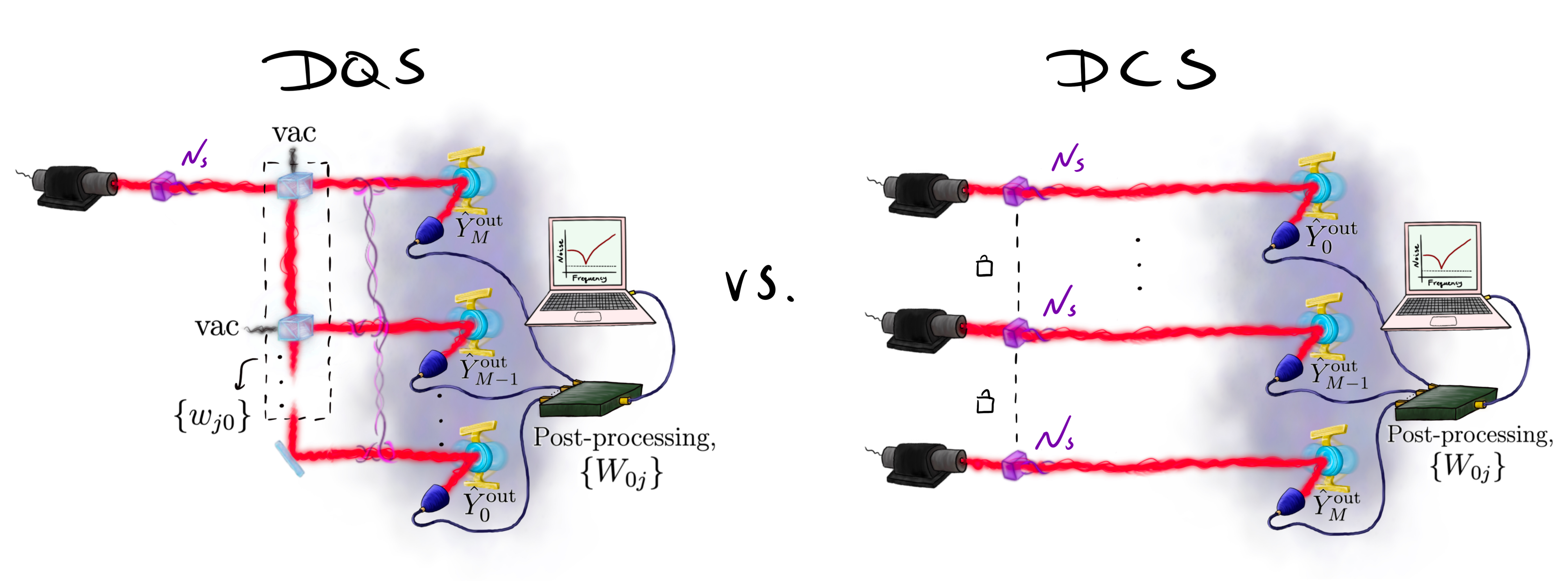}
    \caption{DQS versus DCS. (Left) A distributed quantum sensing (DQS) scheme, where a squeezed vacuum of $N_s$ photons is distributed through a passive linear network---thus generating an entangled state---to an array of $M$ mechanical oscillators. The laser power is $ME_0^2$. (Right) A distributed classical sensing (DCS) scheme with squeezing, where $M$ squeezed vacuum---each with $N_s$ squeezed photons---impinge on an array of $M$ mechanical oscillators. The power per laser is $E_0^2$ (i.e., the total power is $ME_0^2$). These two setups have equal performance, but the DQS scheme consumes only $N_s/M$ squeezed photons per sensor while the DCS scheme consumes $N_s$ squeezed photons per sensor.}
    \label{fig:dqs_dcs}
\end{figure*}
Here, we briefly elaborate on the differences between distributed quantum sensing (DQS) and distributed classical sensing (DCS) schemes; see Fig.~\ref{fig:dqs_dcs}. In the DQS scheme (Fig.~\ref{fig:dqs_dcs}, left), an input entangled state of $M$ modes is prepared by splitting a squeezed vacuum state of $N_s$ photons between the modes. The entangled radiation is then distributed to an array of $M$ mechanical oscillators, after which joint post-processing on the signals occurs. In the DCS scheme with squeezed light (Fig.~\ref{fig:dqs_dcs}, right), $M$ squeezed vacuum---each with $N_s$ number of photons---are generated and independently distributed to $M$ mechanical oscillators, after which joint post-processing on the signals occurs. These two schemes are equivalent in terms of their performance (quantified via, e.g., the SNR or the integrated sensitivity), however the former DQS scheme demands only $N_s/M$ photons per sensor, while the latter DCS scheme demands $N_s$ squeezed photons per sensor. The essential difference between these two setups is that the former utilizes CVMP entanglement to correlate the shot-noise and radiation pressure fluctuations across the sensor array to alleviate the total noise of the mechanics.

\section{Details on experimental projections}
\label{app:exp}

As a concrete example of entanglement-enhanced readout of DM-detectors, we consider the detector proposed in \cite{dal2021VDM}: a membrane-based optomechanical accelerometer deployed as a resonant sensor for vector ultralight dark matter (UDM), specifically, B$-$L UDM. Vector UDM---a field-like DM-candidate expected to produce material-dependent acceleration signals oscillating at the DM Compton frequency---could produce center-of-mass motion in membranes by driving their highly sensitive flexural modes. Cavity-enhanced readout of membrane motion can be further boosted by using a DQS scheme over an array of membrane accelerometers.      

The UDM accelerometers proposed in \cite{dal2021VDM} employ cm-scale stoichiometric silicon nitride (Si$_3$N$_4$) nanomembranes as test masses. Their fundamental flexural modes resonate at 1-10 kHz frequencies, corresponding to 1-100 peV DM mass. Higher-order flexural modes acting as independent test masses and spanning decades of frequency up to $\sim1$ MHz are simultaneously accessible, though we analyze only the fundamental mode here. On resonance, acceleration signals would be amplified by the mechanical quality ($Q$) factor; $Q$'s exceeding 1 billion are achievable in membrane-based silicon nitride structures~\cite{tsaturyan2017ultracoherent,ghadimi2018elastic}. For readout, the membranes would serve as highly reflective (after photonic patterning~\cite{moura2018centimeter,chen2017high}) end-mirrors in optical cavities.

To create Fig.~\ref{fig:dm_projection} of the main text, we consider the detector to be a mg-scale square membrane with a side length of 20 cm and a thickness of 200 nm, resulting in a fundamental resonance frequency around $2$ kHz, whose motion is monitored with an averaging time of 1 year. Following Ref.~\cite{dal2021VDM}, the membrane is assumed to be fixed to a beryllium substrate in order to gain sensitivity to the material-dependent acceleration produced by vector UDM. At an operating temperature of 10 mK and a mechanical quality factor of $Q=10^9$, this system can achieve a thermal-noise-equivalent acceleration resolution of $10^{-12}$ ms$^{-2}/\sqrt{\rm Hz}$ (see Fig.~\ref{fig:noise_squeezed}), corresponding to a minimum detectable DM coupling strength of $g_\text{B-L}= $ $4\times 10^{-25}$. In this example, the membrane serves as an end-mirror in an optical cavity of length $L = 1$ mm with finesse $\mathcal{F}= \pi c / L \kappa = 1000$. Optical readout of the membrane's displacement would be performed using a laser with wavelength $\lambda = 1$ $\mu$m and input power 2 mW, resulting in a shot-noise-limited displacement sensitivity of $9\times 10^{-19}$ m$/\sqrt{\rm Hz}$. 

As depicted in Fig.~\ref{fig:dm_projection} of the main text, a mechanical resonator achieves the best acceleration sensitivity at its resonance frequency. While shot noise limits the detector's off-resonance sensitivity, the dominant noise source on resonance is expected to be radiation pressure backaction at $P=2$ mW input power, with an acceleration noise floor of $2\times 10^{-11}$ ms$^{-2}/\sqrt{\text{Hz}}$ ($g_\text{B-L}= $ $7\times 10^{-24}$). The figure includes additional curves for both classical and DQS schemes, highlighting the potential improvement that can be attained by using multiple ($M=10$) sensors and an entangled light source. The SQL of a classical array (blue dotted) is included, where the laser power is tuned at each frequency to minimize the optical measurement noise. A similar limit is plotted for a DQS setup (red dotted)---where a squeezed vacuum state is distributed uniformly across the sensor array---illustrating the best achievable sensitivity with a fixed squeezing of 10 dB.

\end{document}